\documentclass[preprint]{jpsj2}
\title{
Quantum Resistive Behaviors in Vortex Liquid Regimes at Finite Temperatures
}

\author{
Ryusuke Ikeda
}

\inst{
Department of Physics, Kyoto University, Kyoto 606-8502
}

\recdate{June 7, 2003}

\abst{
   Motivated by a {\it mean field-like} resistive behavior in magnetic fields 
commonly seen in various superconducting cuprates and organics with {\it strong} fluctuation, superconducting (SC) {\it quantum} fluctuation effects on resistive behaviors are reexamined by putting emphasis on their roles in the so-called {\it thermal} vortex liquid regime. By incorporating the quantum fluctuations and a vortex pinning effect in the Ginzburg-Landau (GL) fluctuation theory, it is found that the resistivity $\rho(T)$-curve sharply drops, with no fan-shaped broadening, at a vortex-glass transition point far below an {\it apparent} upper critical field $H_{c2}^*(T)$ as a result of a quantum fluctuation enhanced by an adequately small condensation energy or by a strong field. Fittings to $\rho$-$T$ data of cuprates and organics are performed by phenomenologically including a SC pseudogap region created by high energy SC fluctuations and possible fluctuations of competing non-SC orders. By examining La$_{2-x}$Sr$_x$CuO$_4$ data over a broad doping range, we obtain such conclusions, consistent with recent experimental results, that the in-plane coherence length of hole-doped cuprates decreases with approaching the underdoped limit even in the presence of fluctuating competing orders and that the condensation energy density $(H_c(0))^2 = 0.5 [\phi_0/(2 \pi \lambda(0) \xi_0)]^2$ is maximal near the optimal doping. Further, the case of disordered quasi 2D films showing the field-tuned superconductor-insulator transition is also examined for comparison and discussed in relation to data reported recently. 
}

\kword{
Quantum Fluctuation, Vortex States, Cuprates, Organic Superconductor
}

\begin{document}
\maketitle

\section{Introduction}

Macroscopic behaviors in high temperature cuprate superconductors in nonzero magnetic fields ($H \neq 0$) have led to a renewal of our knowledge on the superconducting (SC) fluctuation. In magnetic fields typically of tesla range applied perpendicular to SC layers, both the resistivity $\rho(T,H)$ and heat capacity in some optimally-doped cuprates show a field-induced fan-shaped broadening behavior near and below an {\it apparent} upper critical field $H_{c2}^*(T)$ to be estimated from thermodynamic quantities. Due to this consistency between $\rho$ and thermodynamic quantities, the resistive broadening was explained as a generic phenomenon in a disordered state (thermal vortex liquid regime) created by the thermal SC fluctuation in nonzero fields\cite{RI1,RI2,Sarti}. After that, the vortex lattice melting transition and its effect on the vanishing of Ohmic resistivity have been studied extensively\cite{RI3,FFH,Natter} as key issues on the vortex phase diagram in real systems. 

However, researches on the vortex phase diagram of cuprates were limited to the case with low enough $H/H_{c2}^*(0)$ values, in which the fluctuation is purely thermal and the vortex pinning effect may be weaker so that the discontinuous nature of the melting transition in pure case may remain intact. The corresponding measurement in higher $H/H_{c2}^*(0)$ values has been realized in overdoped cuprates under several teslas where effects of the pure vortex-solidification are rarely seen. These materials seem to have a longer $T=0$ in-plane coherence length $\xi_0$, and their $\rho$-$T$ curves have shown an apparently mean-field like sharp drop \cite{Mac,HS} particularly in higher fields. Through a comparison with heat capacity data \cite{Carrington}, however, it is clear at present that the sharp resistivity drop has occurred much below $H_{c2}^*(T)$ (or equivalently, $T_{c2}^*(H)$), possibly except at low enough temperatures, suggesting an enhanced fluctuation effect creating a broad vortex liquid regime in those materials. The data in refs.7 and 9 may be understood by merely assuming a thermal SC fluctuation and noting that the normal conductivity $\sigma_n$ in the overdoped Tl-compounds \cite{Mac} is of the order $ 10^2 (R_Q d)^{-1}$: This value is much larger than a typical one of the vortex flow conductivity, i.e., the mean field expression of the superconducting part 
\begin{equation}
\sigma_s=\sigma-\sigma_n
\label{1}
\end{equation}
of the total conductivity $\sigma$, so that $\sigma \simeq \sigma_n$ (the normal part of $\sigma$) even much 
below $H_{c2}^*(T)$, where $R_Q=\pi \hbar/2 e^2=6.45$ (k$\Omega$) is the quantum resistance, and $d$ the distance between the superconducting layers in a layered superconductor. 
However, such a sharp drop of resistivity much below $H_{c2}^*(T)$ accompanied by a diminishing broadening with increasing field has been also observed in other cuprate superconductors \cite{HS,Naito,Mac2,Mac3,Gan,Karpinska,Vedeneev,Capan1,Wang2} and organic superconductors \cite{Sasaki,Ito} with low $H_{c2}^*(T)$ but with much lower $\sigma_n$ values ($< 10 (R_Q d)^{-1}$) and remains unexplained. Equation (1) implies that 
$\sigma_s$ itself is extremely small in the thermal vortex liquid region 
in the tesla range of these materials. 

In this paper, a theory is presented to comprehensively understand such anomalous resistive behaviors in cuprate and organic superconductors with lower $H_{c2}^*(0)$. It is an extension of the previous work \cite{RI1,RI2} to the case with low condensation energy, in which the {\it quantum} SC fluctuation is not negligible in $\sigma_s$ in $H \neq 0$ but induces a resistivity curve following the normal (or, quasiparticle) resistivity curve $\rho_n(T)=\sigma_n^{-1}$ even below $H_{c2}^*(T)$. One of our purposes in this paper is to explain how to evaluate intrinsic material parameters of SC materials with strong fluctuation through fittings to resistivity data. Fitting results to data \cite{Capan1} of underdoped La$_{2-x}$Sr$_x$CuO$_4$ (LSCO) based on the present theory were reported 
in ref.20. 

Here, let us briefly explain why the quantum SC fluctuation in the case is important even at {\it high} temperatures. To do this, it should be first stressed that, in a fixed $H$, a longer $\xi_0$ does {\it not} necessarily imply a weaker fluctuation effect if recalling the fact that the Ginzburg-Landau (GL) fluctuation strength $g_2(H)$ in two dimensional (2D) systems near the zero field transition temperature $T_c$ is 
given by \cite{RI1,Natter}
\begin{eqnarray}
g_2(H, T_c)=\frac{16 \pi^2 \lambda^2(0) k_{\rm B} T_c}{\phi_0^2 d} 
\frac{H}{H_0}
\\ \nonumber 
\propto T_c (\lambda^2(0) \xi_0^2) H,
\label{2}
\end{eqnarray}
where $\phi_0$ is the flux quantum, $\lambda(0)$ is the magnetic penetration depth in $T \to 0$ limit (defined by extraporating from the GL region), and $H_0 \equiv \phi_0/(2 \pi \xi_0^2)$ is the {\it mean field} upper critical field at $T=0$ defined at the microscopic level. A relation between $H_0$ and $H_{c2}^*(0)$ will be given in the following sections. According to eq.(2), an increase of $\xi_0$ under fixed values of other material parameters suggests an enhanced fluctuation at a fixed $H \neq 0$. Note that, in the ordinary 2D GL theory, the Ginzburg number in $H=0$ \cite{FFH} is independent of $\xi_0$ and given by $T_c [\lambda(0)]^2/(\phi_0^2 d)$, except a constant prefactor, which in cuprates decreases monotonically with overdoping. As emphasized elsewhere \cite{RI4}, an increase of $g_2$ leads to an enhancement not only of the thermal fluctuation but of the {\it quantum} one. In contrast to the $H=0$ case, the SC fluctuation in $H \neq 0$ remains massive \cite{RI4} even deep in the vortex liquid region, and thus, its quantum contribution may play an essential role there. In Fig.1, roles of quantum SC fluctuation in the resistivity are illustrated (Details of calculation leading to Fig.1 will be explained later). Figure 1 (a) also includes a comparison with optimally-doped YBCO data \cite{Huebner}. If the {\it quantum} SC fluctuation is taken into account in addition to the thermal one, as the curves in Fig.1 show, an increase of $g_2$ (in this case, of $\lambda(0)$) results in a more sharp drop of $\rho$-$T$ curves just above a 3D vortex-glass (VG) transition \cite{FFH,RI3} lying much below $T_{c2}^*(H)$. Throughout this paper, a filled circle on each $\rho(T)$-curve indicates $T_{c2}^*(H)$. The sharp drop of $\rho$ in Fig.1 (b) is a contrast to the fan-shaped broadening around $H_{c2}^*$ in the case Fig.1 (a) dominated by the thermal SC fluctuation \cite{RI1,FFH}  
and is a consequence of a combination of a (pinning-induced) 3D vortex glass transition and a {\it reduction} of $\sigma_s$ brought by the quantum fluctuations in the {\it thermal} vortex liquid regime. 
\begin{figure}[t]
\scalebox{0.4}[0.4]{\includegraphics{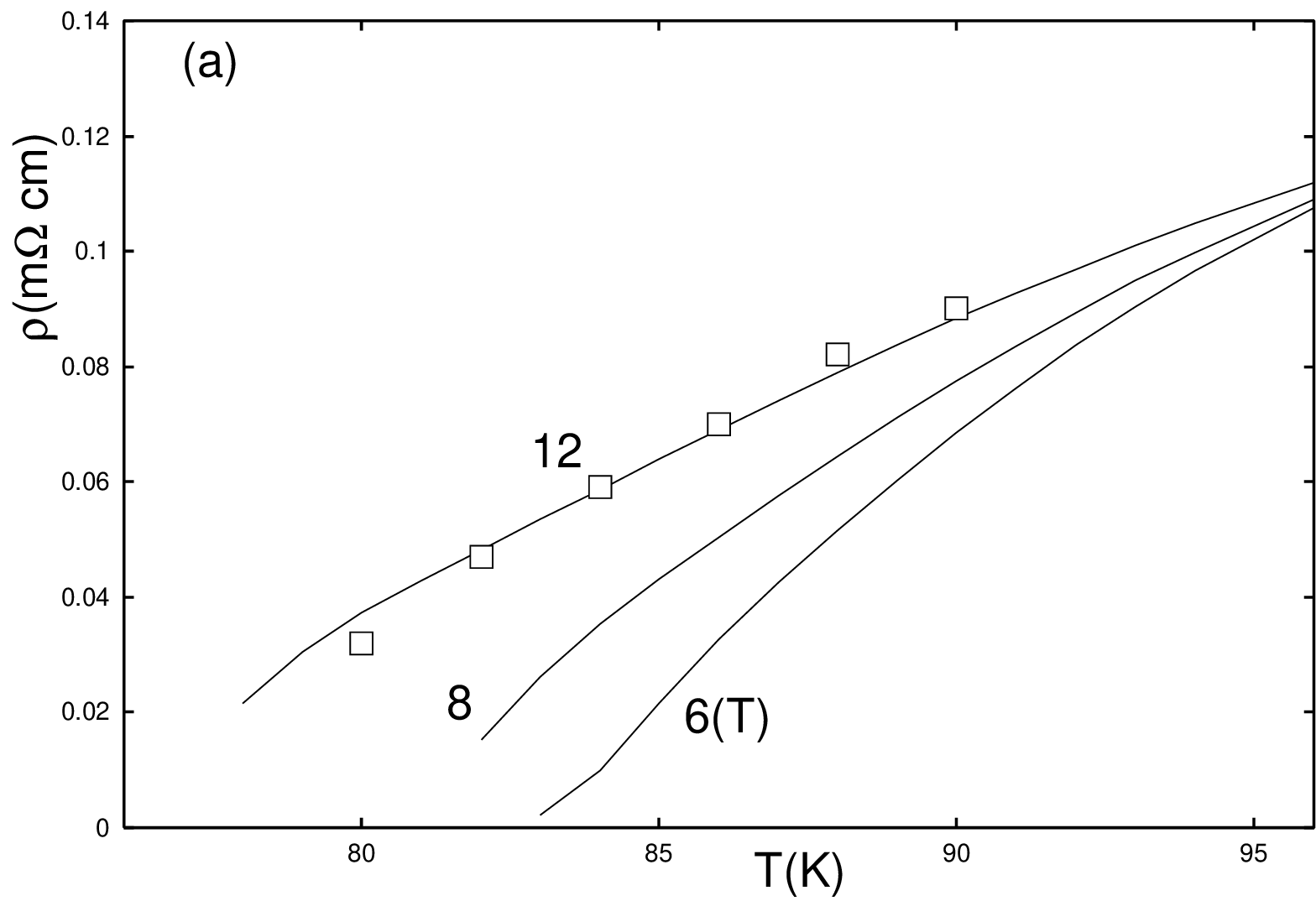}}
\scalebox{0.4}[0.4]{\includegraphics{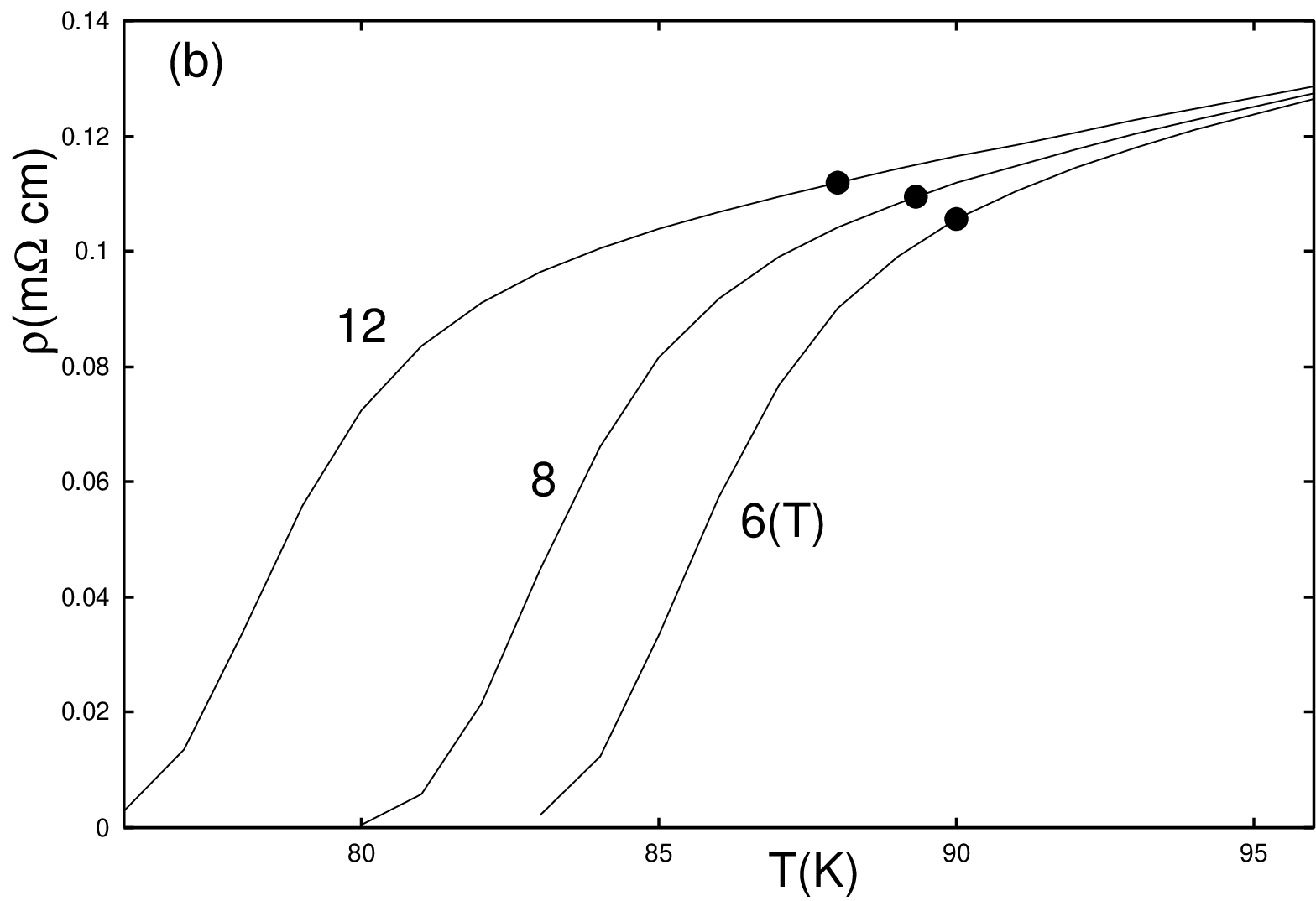}}
\caption{(a) Resistivity curves (solid lines) in 4, 8, and 12(T) calculated in terms of $\lambda(0)=0.11$($\mu$m) and 12 (T) data (symbols) in twinned optimally-doped YBCO \cite{Huebner}. See $\S 3$ for the details of calculations and other parameter values used here. (b) Same curves as those in (a) except the use of $\lambda(0)=0.35$($\mu$m). 
Each darked circle on a $\rho(T)$ curve denotes $T_{c2}^*(H)$ in each $H$.
}
\end{figure}

It has been often argued so far \cite{FFH} in relation to the fluctuation effects in cuprates that the SC fluctuation in bulk materials with a large Ginzburg number near $T_c$ and in $H \neq 0$ will be well described as the thermal fluctuation in the 3D XY model. Based on the above-mentioned fact, this conventional picture is invalid. The SC fluctuation in the limit of strong fluctuation is dominated in $H \neq 0$ by its quantum fluctuation contribution. Then, the resistance follows the normal resistance curve even below $H_{c2}^*(T)$ {\it without} the fan-shaped broadening and shows a mean field-like sudden drop at a VG transition induced by a vortex pinning much below $H_{c2}^*(T)$. 

The above-mentioned reduction of $\sigma_s$ in the thermal vortex liquid 
regime is due to the same origin as the insulating $\sigma_s(T)$ in the {\it quantum} vortex liquid regime near $T=0$ \cite{RI4,RI5}. However, in the temperature range where a pinning-induced VG fluctuation effect is negligible, a nearly classical (vortex flow-like) behavior intervenes these two quantum behaviors so that they can be conveniently seen as independent ones. In fact, the latter, i.e., a reduction of $\sigma_s$ near $T=0$, is limited in most cases to a very low temperature ($T/T_c \ll 1$) window (see Fig.11 below) and is essential to understanding the 
field-tuned superconductor-insulator transition (FSIT) behavior \cite{RI5} which cannot occur without quantum SC fluctuation. Further, thermodynamic quantities such as the magnetization at such low $T$ rapidly vary near $H_{c2}^*(0)$ with sweeping $H$, reflecting a rise \cite{RI4} of dimensionality of fluctuation on approaching $T=0$. In contrast, the reduction of $\sigma_s$, i.e., the flatterning of resistance, in the thermal regime appears, as the example of Fig.1 shows, even at high temperatures comparable with $T_c$ in systems with moderately strong fluctuation, and the corresponding $T$-dependence of thermodynamic quantities is broadened as the quantum fluctuation is stronger (see Figs.8 and 9 below). 

Roles of quantum fluctuation of vortex positions in a perfectly clean vortex 
solid have been examined by Blatter and Ivlev \cite{Blatter} as an explanation of high field behaviors of the first order melting line of a clean vortex solid in optimally-doped YBCO. Note that the fluctuation of vortex positions is included in SC fluctuations because the vortex positions are nodes (zero points) of the SC order parameter. However, it is well understood that the quantum effect on the melting transition of the optimally-doped samples is usually negligible. For instance, an observed field-induced deviation (reduction) of the melting line from the low field behavior is quite small and can be understood rather as a consequence of pinning disorder which is more effective with increasing fields \cite{RI3}. On the other hand, the quantum fluctuation effect in the thermal vortex liquid regime has not been examined there \cite{Blatter}. Since the fluctuation with lower energy becomes more important upon cooling in the thermal vortex liquid regime (i.e., the disordered non SC phase), it is clear that the quantum fluctuation is much more important, e.g., near $H_{c2}^*(T)$ rather than near the 
melting line. As Fig.1 (a) shows, however, the quantum effect on the resistivity curves, and hence, on the melting transition line, of optimally-doped YBCO 
is quite negligible. When examining in this paper resistivity curves suggestive of a remarkable quantum fluctuation effect, a corresponding quantum effect on a VG transition line replacing the quantum melting line \cite{RI4} in clean limit is relatively negligible and will not be examined 
theoretically. 

This paper is organized as follows. First, in $\S 2$, a semi-quantitative theory describing transport phenomena in the thermal and quantum vortex liquid regimes of 2D-like systems is given. Results in $\S 2$ are used in the ensuing 
sections by incorporating microscopic details to examine data on superconducting cuprates \cite{RI7} ($\S 3$) and organics \cite{RI6} ($\S 4$), and it is shown that the rapid vanishing of resistivity, often seen in these materials, is a consequence of a competition between a strong quantum SC fluctuation and a 3D VG fluctuation. For comparison, resistivity curves in disordered quasi 2D systems with $s$-wave pairing are also discussed in $\S 5$, and it is emphasized through Fig.13 that the rapid vanishing of resistivity much below $H_{c2}^*(T)$ cannot be peculiar to the strongly-correlated electron systems. 
In addition to data analysis in nonzero fields, effects of the SC pseudogap width and the quantum fluctuation on the critical region near $T_c$ in $H=0$ are also considered in $\S 6$ and an appendix. These two ingredients are expected to be the origins of the unexpectedly \cite{Wang1} narrow $H=0$ critical region in underdoped cuprates. 

In contrast to the dc electric conductivities and the Nernst coefficient, any SC fluctuation effect in a nonsuperconducting (non-SC) phase does not appear, at least in the GL approach, in the thermal conductivity \cite{Smith}. The latter type of quantities are dominated by a quasiparticle transport \cite{Tesanovic} even much below $H_0$. In this paper, we focus primarily on the SC contributions of the former ones and assume their quasiparticle contributions (such as $\sigma_n$) to be estimated from experimental data. Since a strong quantum fluctuation deviates $\sigma_s$ from its mean field vortex flow expression $\sigma_{\rm MF}$, we do not expect a possible change of $\sigma_{\rm MF}$ arising from the gap nodes in unconventional superconductors to lead to a serious discrepancy in our results. Actually, the resistivity defined from the microwave surface impedance data \cite{Matsuda} is usually remarkably different from dc resistivity data and is rather comparable with $\sigma_{\rm MF}$ derived on a single vortex level. 

\section{Expression of Transport Quantities}

In this section, theoretical expressions useful in examining and fitting experimental data of the resistivity $\rho$ and the Nernst coefficient $N$ (or, the transport entropy $s_\phi$) are derived. Although the basic framework of the theory is essentially the same as that given in ref.23, we need here to reexpress it in a form applicable to superconductors other than the $s$-wave dirty films considered there. Our method of renormalization of the SC fluctuation is essentially an extension of the Hartree approximation \cite{RI1,UD} to the case with quantum fluctuation and a vortex pinning effect. 
Readers who are not interested in the details of derivation of the theoretical results may skip the main part of this section and jump to the final paragraph of this section. 

We start with a 2D GL action 
\begin{eqnarray}
S &=& d \int d^2r \biggl[  \beta \sum_\omega (\psi_\omega({\bf r}))^* \gamma({\bf Q}^2)|\omega| \psi_\omega({\bf r}) 
\\ \nonumber \cr
&+& \int_0^\beta \! d\tau \biggl( u({\bf r}) |\psi({\bf r}, \tau)|^2 
+ (\psi({\bf r}, \tau))^* \mu({\bf Q}^2) \psi({\bf r}, \tau) 
+ {b \over 2} |\psi({\bf r}, \tau)|^4  \biggr) \biggr],
\label{3}
\end{eqnarray} 
as a model of a 2D-like layered superconductor under a field ${\bf H}$ perpendicular to the layers. Here, ${\bf Q} = -i \nabla + (2 \pi/\phi_0) {\bf A}({\bf r})$ is the 2D gauge-invariant gradient, $\psi(\tau) = \sum_\omega \psi_\omega e^{-i\omega\tau}$ is a single componet pair-field (SC order parameter), $\beta$ the inverse temperature, $\omega=2 \pi n/\beta$ with integer $n$, $\tau$ the imaginary time, and $b > 0$. The random potential $u({\bf r})$ has zero mean and satisfies ${\overline {u({\bf r}) \, u({\bf r'})}} = b_p({\bf r}-{\bf r}')$. Although the nonlocality of $b_p({\bf r})$ is not negligible in $T \to 0$ limit \cite{RI5}, we assume that $b_p({\bf r})$ can be replaced hereafter by $\delta({\bf r})$ multiplied by a coefficient $b_p$ because no fluctuation effects close to zero temperature are considered in this paper. Further, the 3D nature due to a coupling between the SC layers was neglected in writing eq.(3) by assuming a strong anisotropy. This assumption is based on the fact found through the previous data fittings \cite{RI1,RI2} that, as far as the vortex pinning is ineffective, the effect of the layer coupling on the {\it in-plane} electric conductivities in 2D-like systems is extremely weak even deep in the liquid regime. Hence, the 3D nature will be incorporated later in considering a VG contribution to the conductivity. An additional dynamical term $i \gamma' \omega [\psi_\omega]^* \psi_\omega$ leading to a fluctuation Hall effect and resulting from a particle-hole assymmetry was neglected in eq.(3). This is justified as far as $|\gamma'| \ll \gamma$. 

When the GL approach is applied to the low $T$ and high $H$ region in which any phase-only approach is inapplicable, $H$-dependences of the coefficients $\gamma$, $\mu$, and $b$ need to be taken into account since the familiar low $T$ divergences \cite{BK} of these coefficients in zero field are cut off by the orbital depairing effect of the magnetic field. Thus, their ${\bf Q}$-dependences should not be treated perturbatively. Although, strictly speaking, the Pauli paramagnetic depairing effect will also suppress the low $T$ divergences, it will be assumed that it becomes important only in much higher fields than the field range which is focused on in the present work. Then, by expanding $\psi$ in terms of the Landau levels (LLs), $\psi({\bf r})=\sum_n \varphi_{n, p} u_{n, p}({\bf r})$, 
the coefficients $\gamma$ and $\mu$ of quadratic terms of eq.(3) are represented as $n$-dependent ones, $\mu_n$ and $\gamma_n$, where $n$ ($\geq 0$) is the LL index, $p$ is a quantum number measuring the degeneracy in each LL, and $u_{np}$ is an eigenfunction in $n$-th LL. 
For instance, the {\it bare} fluctuation propagator ${\cal G}_n^{(0)}(|\omega|) = <|\varphi_{n, p}(\omega)|^2>$ valid in the $b$, $b_p \to 0$ limit is expressed by ${\cal G}_n^{(0)}(|\omega|) = (\gamma_n |\omega| + \mu_n)^{-1}$, where the triangular bracket denotes the ensemble average on $\psi$. The microscopic mean field transition point $T_0(H)$ (or $H_0(T)$) is defined by $\mu_0=0$. Hereafter, $T_0(0)$ is often written merely as $T_0$. Detailed forms of $\gamma_n$ and $\mu_n$ will be given separately in the following sections. On the other hand, no $n$-dependence of the coefficients $b$ and $b_p$ need to be specified throughout this paper, since a high $H$ approximation is invoked below in obtaining expressions useful for analyzing experimental data, and hence, the mutual interaction between the SC fluctuations and the interaction between a fluctuation and the random potential $u$ are considered primarily within the lowest LL (LLL) with $n=0$. In particular, in examining the data of cuprates and organics in which microscopic descriptions are still controversial, it is appropriate to assume that $b$ is one of material parameters related to the magnetic penetration depth $\lambda(0)$, while the pinning strength $b_p$ is an extrinsic parameter dependent on real samples used in experiments. 

\begin{figure}[t]
\scalebox{0.4}[0.4]{\includegraphics{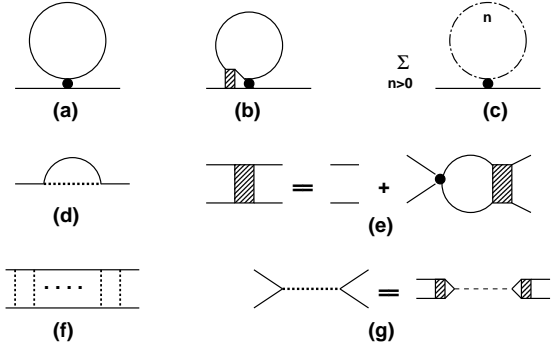}}
\caption{Diagrams expressing the four terms of r.h.s. of eq.(4) ((a) to (d)) and the VG susceptibility (f). The interaction vertex correction (the hatched rectangle) is expressed in the diagram (e), and the diagram (g) defines ${\tilde b}_p$. The solid lines, the solid dots, the chain line, and the thin dashed line in (g) express, respectively, ${\cal G}_0$, $b$, ${\cal G}_n$ ($n \geq 1$), and $b_p$.}
\end{figure}

To renormalize the $\psi$-fluctuation with low energy, 
the LLL approximation will be 
used. That is, a relatively high field range is assumed, in which the renormalized energy scales (masses) of higher ($n \geq 1$) LL fluctuations are much larger than that of LLL and do not deviate remarkably from their bare ones $\mu_n$ less sensitive to $T$ compared with $\mu_0$. Then, the renormalization of higher LL modes can be assumed to have already been accomplished independently. Further, the vortex pinning effect in the LLL-fluctuation renormalization will be incorporated to the one loop order \cite{RI5}. Then, the {\it renormalized} LLL fluctuation propagator ${\cal G}_0(|\omega|)$ is determined by 
\begin{eqnarray}
({\cal G}_0(|\omega|))^{-1} - \gamma_0|\omega| - \mu_0 
= \Sigma_0 
+ \Delta \Sigma_{\rm l} \;
\\ \nonumber
+ \Delta \Sigma_{\rm h} 
- \frac{{\tilde b}_p}{2 \pi r_B^2 d} {\cal G}_0(|\omega|). 
\label{4}
\end{eqnarray}
Diagrams for the four terms of the r.h.s. of eq.(4) are given 
by Fig.2 (a) to (d). Here, the factor ${\tilde b}_p$ is a renormalized pinning 
vertex, sketched in Fig.2 (g), 
and is given by 
\begin{equation}
{\tilde b}_p = b_p \int \frac{d^2k}{2 \pi}  
e^{-{\bf k}^2/2} (1 + u_{\rm v} 
e^{-{\bf k}^2/2})^{-2} = \frac{b_p}{1+u_{\rm v}},
\label{5}
\end{equation}
where $u_{\rm v} = b \beta^{-1} \sum_\omega 
{\cal G}_0^2(|\omega|)/(2 \pi r_B^2 d)$. Consistently with this approximation, the VG transitioin point defined in the Gaussian approximation of VG fluctuation (i.e., the mean field VG transition point) is determined according to the ladder diagram Fig.2 (f) expressing the VG susceptibility \cite{RI5} 
as the limit $\mu_{{\rm vg}, 0} \to +0$, where 
\begin{equation}
\mu_{{\rm vg},0}= 1 - \frac{{\tilde b}_p [{\cal G}_0(0)]^2}{2 \pi r_B^2 d}. 
\label{6}
\end{equation}
Although, strictly speaking, this pinning-renormalization replacing $b_p$ by ${\tilde b}_p$ is merely valid far above the vortex-solidification line in the pinning-free limit \cite{RI3} and is diagrammatically consistent just with the pinning-free fluctuation renormalization in the case with no $\Delta \Sigma_l$ (see below regarding $\Delta \Sigma_l$), this approximation will be used hereafter for practical purposes because all resistivity data we will examine below belong to the cases where the first order vortex solid-liquid 
transition was destroyed by the pinning. 

The main roles of LLL mass renormalization are played by the Hartree 
term $\Sigma_0$ of the self energy corresponding to Fig.2 (a) 
and expressed as 
\begin{equation}
\Sigma_0 = \frac{b}{2 \pi r_B^2 d \beta} \sum_\omega {\cal G}_0(|\omega|). 
\label{7}
\end{equation}
For the coefficient $b$, the ordinary GL expression \cite{RI1} $16 \pi^2 \lambda(0)^2/(\phi_0 H_0)$ in low $H/H_0$ will be used near and below $T_{c2}^*(H)$. The validity of this identification will be discussed in $\S 6$ in relation to the cuprates.  
Regarding the additional renormalization (correction) term $\Delta \Sigma_l$ within the LLL, an approximation of RPA type sketched in Fig.2 (b) 
\begin{equation} 
\Delta \Sigma_l= \beta^{-1} \sum_\omega {\cal G}_0(|\omega|) \, \frac{{\rm ln}[1 + b E_{00}(|\omega|)/(\pi r_B^2 d)]}{2 E_{00}(|\omega|)}, 
\label{8}
\end{equation}
(see eq.(2.11) of ref.21) will be useful below, where $E_{00}(|\Omega|)=\beta^{-1} \sum_{\omega} {\cal G}_0(|\omega|) {\cal G}_0(|\omega+\Omega|)$. The term $\Delta \Sigma_l$ is negligible in the thermal 2D case \cite{RI1} and, as far as a qualitative study of resistivity curves is concerned, may be negligible even in the quantum 2D case. However, to make sure, this term will be included when attempting to fit resistivity data in $\S 3$ and $4$. 
\begin{figure}[t]
\scalebox{0.5}[0.5]{\includegraphics{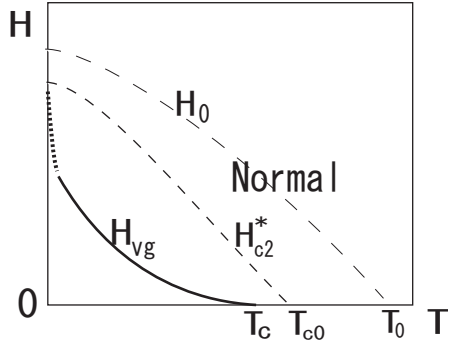}}
\caption{Schematic $H$-$T$ diagram in which various crossover or transition curves defined in the text are drawn. The low $H$ limit of $H_{\rm vg}(T)$-line corresponds to $T_c$.} 
\label{} \end{figure}

The term $\Delta \Sigma_h$ implies higher LL contributions to the LLL mass renormalization. It should be remarked \cite{RI95} that the higher LL fluctuations are not negligible even in the present high $H$ approximation but do contribute to a downward shift of $T_0(H)$ to $T_{c2}^*(H)$ corresponding to the apparent upper critical field $H_{c2}^*(T)$ , just like in zero field case \cite{HN} where a reduction $\Delta T_c$ of the mean field transition point is ascribed to the amplitude-dominated fluctuation with {\it high} energy. This downward shift is conveniently represented, as in Fig.2 (c), by a sum of Hartree diagrams. The resulting temperature $T_0 - \Delta T_c$ corresponds not to the true SC transition point $T_c$ at which the linear resistance vanishes but, roughly speaking, to an onset of a sharp resistive vanishing or of the critical region accompanying the transition at $T_c$. Rather than $T_c$, this resistive onset, denoted as $T_{c0}$ hereafter, appears in the ensuing expressions for $H \neq 0$. Since 
Fig.2 (c) is ultraviolet divergent even in 2D case when the quantum fluctuation is included, its $H$-dependence is a correction, and hence, $\Delta \Sigma_h$ may be identified with its expression in $H=0$ case. Further, when the presence of other non-SC order parameter fluctuations coupling to the SC fluctuation is not negligible as in the underdoped cuprates ($\S 3$), they should also be incorporated into $\Delta \Sigma_h$. It will be argued in $\S 3$ that even such non-SC fluctuations do not bring additional $H$-dependences. Then, $\Delta \Sigma_h$ may be written as ${\rm ln}(T_0/T_{c0})$ \cite{RI4,RI95}, and the $T_{c2}^*(H)$-line, reducing to $T_{c0}$ in $H \to 0$ limit, is determined by $\mu_0 + \Delta \Sigma_h =0$. In Fig.3, the characteristic temperatures and fields defined above are sketched in the $H$-$T$ diagram. 

Due to the presence of the pinning term $\propto {\tilde b}_p$, solving eq.(4) selfconsistently requires not a difficult but a very cumbersome numerical-integration even if neglecting $\Delta \Sigma_l$. A more cumbersome situation is encountered \cite{RI5} when trying to present a practical expression on the VG contribution $\sigma_{\rm vg}$ to the conductivity. Since providing theoretical formulas useful in analyzing experimental data is the main purpose in this section, we will not try to give a selfconsistent solution of eq.(4) but invoke the following approximation to give practically convenient expressions: The form ${\cal G}_0(|\omega|)=(\, \gamma_0|\omega| + ({\cal G}_0(0))^{-1} \, )^{-1}$ will be assumed. That is, any pinning-induced renormalization of frequency dependences will be neglected. Then, eq.(4) is replaced by  
\begin{equation}
({\cal G}_0(0))^{-1} = \mu_0+{\rm ln}(T_0/T_{c0}) + \Sigma_0 + \Delta \Sigma_l - \frac{{\tilde b}_p}{2 \pi r_B^2 d} {\cal G}_0(0).
\label{9}
\end{equation}
Consistently with this, $\Sigma_0$ is represented as 
\begin{equation}
\Sigma_0= \frac{b}{2 \pi^2 r_B^2 d \gamma_0}  \int_0^{\epsilon_c} d\epsilon \ {\rm coth}\biggl(\frac{\beta \epsilon}{2 \gamma_0} \biggr) \ 
\frac{\epsilon}{\epsilon^2+({\cal G}_0(0))^{-2}} 
\label{10}, 
\end{equation}
where the cutoff $\epsilon_c$ is a constant of order unity, 
and $E_{00}(\omega)$ becomes 

\begin{eqnarray}
E_{00}(|\Omega|)= \beta^{-1} \sum_\omega {\cal G}_0(|\omega|+|\Omega|) \, {\cal G}_0(|\omega|) 
\\ \nonumber
\times \biggl(1 + \frac{\gamma_0 |\Omega|}{2} {\cal G}_0(|\Omega|/2) \biggr)
\label{11}
\end{eqnarray}
after simply arranging the $\omega$-summation. 
Although, strictly speaking, this simplification is valid when $1-\mu_{g,0} \ll 1$, its justification should be discussed rather through a fitting to experimental data. Through computations leading to the results in the ensuing sections, we verified that refining this approximation was unnecessary except in $s$-wave disordered films at low enough temperatures (see $\S 5$). 

Next, bearing eq.(9) describing the SC fluctuation in LLL in mind, expressions of transport quantities will be derived in terms of Kubo formula for them. 
The superconducting part $\sigma_s$ of the 
electric (diagonal) conductivity is written in the form 
\begin{equation} 
d \ \sigma_s = \biggl(- {\partial \over {\partial \Omega}} \biggr) \ {\overline {\langle j_{e, \ x}(-{\rm i}\Omega) \ j_{e, \ x}({\rm i}\Omega) \rangle}} \biggl|_{\Omega \to +0},
\label{12}
\end{equation}
where the overbar (bracket $\langle \,\,\, \rangle$) implies the random ($\psi$) average. The corresponding expression 
of the transport entropy $s_\phi$ is naively assumed to take the form \cite{Dorsey} 
\begin{equation}
d \ s_\phi = \beta 
\phi_0 \biggl(- {\partial \over {\partial \Omega}} \biggr) \ 
{\overline {\langle j_{e, \ x}(-{\rm i} \Omega) \ j_{h, \ y}({\rm i}\Omega) \rangle}} \biggl|_{\Omega \to +0}
\label{13}
\end{equation}
(see, however, the next paragraph). 
If using the real time representation, the spatially-averaged electric and heat currents for the ordinary phenomenological GL model are given by 
\begin{eqnarray}
j_{e, \ x} &=& \xi_0^2 \frac{2 \pi}{\phi_0} \langle \psi^* Q_x \psi 
+ {\rm c}. \ {\rm c} \rangle_{\rm sp}. \\ \nonumber 
j_{h, \ y} &=& \xi_0^2 \langle {\rm i} [\partial_t \psi^*(t)] Q_y \psi(t) + {\rm c}. \ {\rm c} \rangle_{\rm sp}, 
\label{14}
\end{eqnarray}
respectively, where $t$ is the real time, and the bracket $\langle \, \, \, \rangle_{\rm sp}$ denotes the spatial average. 
Within the high $H$ approximation in which ${\bf j}_e$ consists of LLL and the next lowest LL, the prefactor $\xi_0^2 (2 \pi/\phi_0)$ of ${\bf j}_e$ in eq.(14) generally becomes $\pi r_B^2/(\phi_0 \, {\cal G}_1(0))$ in the vortex liquid region {\it irrespective of the microscopic details}. This fact proved in ref.23 (see eq.(2.23) there) is a direct consequence of gauge-invariance. 

In contrast, the prefactor of ${\bf j}_h$ may be, at least at low $T/T_0$, strongly affected by a microscopic mechanism independent of the fluctuation property, and the correct $s_\phi$ should vanish in $T/T_0 \to 0$ limit. This issue in the $s$-wave case can be seen in ref.35. Since a correct magnitude of the prefactor of $s_\phi$ is unimportant for the purpose in this paper, for convenience, eq.(13) is used hereafter to obtain the transport energy $U_\phi$. However, some comments on this point will be necessary because, strictly speaking, a {\it direct} use itself of eq.(13) in obtaining a correct $U_\phi$ is not justified. Based on microscopic and thermodynamic results \cite{Maki}, the factor of $j_{h,y}$ in eq.(14) has to be doubled near $T_{c0}$, and a contribution $- \phi_0 M$ from the magnetization current has to be subtracted from {\it the resulting} $T s_\phi$, where $M$ ($< 0$) is the SC magnetization. As is well known, however, $M$ in high fields, where the LLL approximation is useful, is always given, except a prefactor, by the mean squared pair-field $<|\psi|^2> = \Sigma_0/b$, while we will explain below that the same thing holds true for $T s_\phi$ given by eq.(13). Consequently, as far as the transport energy $U_\phi=T s_\phi$ in LLL is concerned, $U_\phi$ defined by eq.(13) with eq.(14) coincides with a microscopically valid one. 

Derivation of the term in $\sigma_s$ with no pinning-induced vertex correction, denoted hereafter as $\sigma_{\rm fl}$, is almost the same as that in the pinning-free case \cite{RI3,RI4}. As already mentioned, the higher LLs are assumed to be inert in the LLL mass-renormalization by invoking a situation in high $H$ or deep in the vortex liquid regime. In fact, reflecting the equivalence between the next lowest LL mode and the compressional elastic mode of the vortex liquid \cite{RI95}, $({\cal G}_1(0))^{-1}$ has to reduce in the vortex liquid regime to $\mu_1-\mu_0$ insensitive to $T$. Further, within the high $H$ approximation, fluctuation vertex corrections accompanied by an interaction between the LLL and the next lowest LL modes can be neglected in the $\sigma_{\rm fl}$-expression. Then, our calculation of $\sigma_{\rm fl}$ is the same as the previous one \cite{RI4}, and one finds 
\begin{eqnarray}
&d R_Q \sigma_{\rm fl}& = {\gamma_0 \over {2 ({\cal G}_1(0))^2 \beta}} \sum_\omega \biggl[ {\cal G}_0(\omega) {\cal G}_1(\omega) 
[ {\cal G}_0(\omega) + g {\cal G}_1(\omega) ] - {{({\cal G}_0(\omega))^2 + g^2 ({\cal G}_1(\omega))^2} \over {({\cal G}_1(0))^{-1} + g ({\cal G}_0(0))^{-1}}} \biggr] 
\\ \nonumber \cr
&=& ({\cal G}_1(0))^{-2} \int \frac{d\epsilon}{2 \pi} \frac{\beta \gamma_0 \gamma_1 \epsilon^2}{2 {\rm sinh}^2(\beta \epsilon/2)} 
\frac{1}{(\gamma_0^2 \epsilon^2 + ({\cal G}_0(0))^{-2})(\gamma_1^2 \epsilon^2 + ({\cal G}_1(0))^{-2})},
\label{15}
\end{eqnarray}
where $g=\gamma_1/\gamma_0$, ${\cal G}_1(\omega)=(\gamma_1|\omega|+({\cal G}_1(0))^{-1})^{-1}$, and the prefactor $({\cal G}_1(0))^{-2}$ arises from the above-mentioned ${\bf j}_e$-vertices. As mentioned in ref.20, however, the details of ${\cal G}_1(0)$ are not reflected in computed results of $\sigma_{\rm fl}$, in most of the $T$ and $H$ ranges we have examined, as a result of the relation $\gamma_1 {\cal G}_0(0)/(\gamma_0 {\cal G}_1(0)) \ll 1$. For this reason, 
\begin{equation}
({\cal G}_1(0))^{-1} = \mu_1-\mu_0
\label{16}
\end{equation}
will be assumed hereafter both above and below $T^*_{c2}(H)$. 
Clearly, $\sigma_{\rm fl}$ of eq.(15) vanishes in $T \to 0$ limit. Note that the neglect based on the high $H$ 
approximation of the fluctuation vertex corrections in $\sigma_{\rm fl}$ 
does {\it not} conflict with the inclusion of $\Delta \Sigma_l$ in the mass renormalization. 

In the realistic case with a vortex pinning, an additional contribution $\sigma_{\rm vg}$ to $\sigma_s$ created by a pinning-induced vertex correction becomes divergent on approaching a 3D VG transition point $T_{\rm vg}$ \cite{FFH,RI8} from above. Near $T_{\rm vg}$, the contribution of quantum SC fluctuation to $\sigma_{\rm vg}$ is negligible, and a $\sigma_{\rm vg}$-expression in the thermal case 
\begin{equation}
d R_Q \sigma_{\rm vg} = c_p \frac{\gamma_0 T_{c0}}{(t-t_{\rm vg})^s},
\label{17}
\end{equation}
where $t=T/T_{c0}$, will be used in comparing our theory with experimental data by choosing $t_{\rm vg}=T_{\rm vg}/T_{c0}$ and the prefactor $c_p$ as being sample-dependent (i.e., pinning-dependent and, in the case of cuprates, doping-dependent) parameters. Although the prefactor $c_p$ should depend not only on $b_p$ and $b$ (i.e., the pinning and fluctuation strengths in $H=0$) but on $H$ \cite{RI8}, for simplicity, it will be assumed to be $H$-independent in $\S 3$ and $4$. The exponent $s$ is known to depend on the dimensionality of pinning potentials dominant in the sample we focus on, and, strictly speaking, it is difficult to predict an appropriate value of $s$ through each fitting far above $T_{\rm vg}$. Throughout the fittings to be explained below, $s=4.0$ was always assumed. 

On the other hand, in situations where the system is 2D-like in spite of the presence of pinning effect, a true divergence of $\sigma_{\rm vg}$ may not occur, and hence the quantum fluctuation contribution to $\sigma_{\rm vg}$ is not negligible. As an approximate $\sigma_{\rm vg}$-expression appropriate to this situation, the expression derived within the Gaussian approximation in ref.23
\begin{eqnarray}
d R_Q \sigma_{\rm vg} &=& (\pi \beta)^{-1} \biggl(\frac{b_p {\cal G}_0(0)}{2 \pi r_B^2 d} \biggr)^2 
\sum_{\omega} {\partial \over {\partial |\omega|}} \biggl[ {\rm ln}(1 + c_c^{-2} \xi_{{\rm vg}, 0}^2) {{{\cal G}_0(\omega)} \over 2} ({\cal G}_0(\omega) + {\cal G}_0(0)) \\ \nonumber \cr
\! &-& \! ({\cal G}_0(\omega))^2 {\rm ln}\biggl({{1 + c_c^{-2} \xi_{{\rm vg}, 0}^2 + 2 |\omega| \gamma_0 {\cal G}_0(0) [\xi_{{\rm vg}, 0}]^4} \over {1 + 2|\omega|\gamma_0 {\cal G}_0(0) [\xi_{{\rm vg},0}]^4}} \biggr) 
\ \biggr]
\label{18}
\end{eqnarray}
will be used below, where $\xi_{{\rm vg},0}=\mu_{\rm vg}^{-1/2}$ is the dimensionless VG correlation length expressed in unit of $r_B$, and $\mu_{\rm vg}=\mu_{{\rm vg},0}+\sqrt{\gamma_0 T {\cal G}_0(0)}$. This form of $\mu_{\rm vg}$ is an expression useful for interpolation near an apparent 2D quantum VG transition field $B_{\rm vg}^*$ \cite{RI5}, and the term $\sqrt{\gamma_0 T {\cal G}_0(0)}$ plays a role of modelling the presence of the quantum VG critical regime within the Gaussian approximation. Further, the constant $c_c$ of order unity is related to an upper cutoff of the wavevector integrals and will be hereafter chosen as $c_c=1$. 

In general, $\sigma_{\rm vg}$ also vanishes in $T \to 0$ limit in $B > B_{\rm vg}^*$, where $B_{\rm vg}^*$ is defined by $\mu_{{\rm vg},0}(B=B_{\rm vg}^*, T \to 0) \to 0$. Further, $\xi_{{\rm vg},0}(B=B_{\rm vg}^*, T) \propto T^{-1/4}$, and $d R_Q \sigma_{\rm vg}(B=B_{\rm vg}^*)$ is approximated by a {\it nonuniversal} \cite{RI5,com1} constant at low enough $T$ where $\xi_{{\rm vg},0} \gg 1$. Note that eq.(18) is an expression valid within the Gaussian approximation and hence, may diverge in $B < B_{\rm vg}^*$ at a finite temperature like in 3D case. As is seen later in Fig.12, however, it is possible that it remains nondivergent at nonzero $T$, depending on the microscopic details. In the fittings, we will use either eqs.(16) 
or (18), depending upon the situations. 

One might wonder if the fact that both of eqs.(15) and (18) in $B > B_{\rm vg}^*$ vanish in low $T$ limit is not a result of the neglect of some vertex corrections in the Kubo formulas. However, it was proved \cite{RI4} at least in the pinning-free case that all terms including the vertex corrections vanish in low $T$ limit. It is trivially performed to extend the proof in ref.21 to the case with vortex-pinning as follows. First, as in the electron systems \cite{PWA}, as far as the conductivity {\it prior to} the random average is considered, the {\it fluctuation} propagators ${\cal G}_n(|\omega|)$ ($n \geq 0$) depend on two coordinates and can be represented in a form like ${\cal G}_n(|\omega|; {\bf r}_1, {\bf r}_2) = \sum_\mu u_\mu({\bf r}_1) {\cal G}_{n, \mu}(|\omega|) u_\mu^*({\bf r}_2)$ where $u_\mu$ is an eigen function specified by a quantum number $\mu$. Since the proof in ref.21 is applied in the same way as far as the spectral form (i.e., frequency dependence) in the low frequency limit remains dissipative and is valid irrespective of the forms of coordinate or wavevector dependences of ${\cal G}_n$, it is concluded that the sum $\sigma_{\rm fl} + \sigma_{\rm vg}$ vanishes in low $T$ limit even if the vertex corrections are included. Therefore, by combining the high $H$ approximation with this, the neglect of the vertex correction in eqs.(15) and (18) is safely valid. 

Just as for the conductivity, the transport energy $U_\phi$ may also be examined by neglecting the vertex corrections in the Kubo formula because the heat current is also accompanied by the next lowest LL mode. On the other hand, the pinning-induced vertex correction related to the VG fluctuation may be neglected in $s_\phi$ assuming a weak pinning because no divergent contribution near $T_{\rm vg}$ will arise in this quantity as a result of the additional time-derivative in the ${\bf j}_h$-expression compared to the ${\bf j}_e$-one. By arranging the frequency summation to take the $\Omega$-derivative, we obtain

\begin{eqnarray}
&s_\phi& = \frac{H} {d H_0(0) {\cal G}_1(0)} \sum_\omega \biggl[ \gamma_0 \gamma_1 |\omega| ( [\gamma_1 {\cal G}_1(0)]^{-1} 
- [\gamma_0 {\cal G}_0(0)]^{-1} ) ({\cal G}_0(\omega) \, 
{\cal G}_1(\omega))^2 
\\ \nonumber \cr
&+& \frac{1}{(\gamma_0 {\cal G}_0(0))^{-1} + (\gamma_1 {\cal G}_1(0))^{-1}} \biggl[ ([\gamma_1 {\cal G}_1(0)]^{-1} - [\gamma_0 {\cal G}_0(0)]^{-1}) {\cal G}_0(\omega) {\cal G}_1(\omega) \\ \nonumber \cr
&-& \frac{\gamma_0}{\gamma_1} |\omega| ({\cal G}_0(\omega))^2 
+ \frac{\gamma_1}{\gamma_0} |\omega| ({\cal G}_1(\omega))^2 \biggr] 
\biggr].
\label{19} 
\end{eqnarray}

In the present high $H$ approximation, eq.(19) is simplified, by neglecting terms of higher order in $\gamma_1 {\cal G}_1(0)/(\gamma_0 {\cal G}_0(0))$, as 
\begin{equation}
U_\phi \simeq \frac{\beta^{-1} H}{H_0 d {\cal G}_1(0)} \sum_\omega {\cal G}_1(\omega) {\cal G}_0(\omega) \simeq \frac{\phi_0^2}{16 \pi^2 \lambda^2(0)} 
\Sigma_0, 
\label{20}
\end{equation}
where, as mentioned earlier, 
the prefactor $({\cal G}_1(0))^{-1}$ is carried by 
the ${\bf j}_e$-vertex. Namely, in the present high $H$ approximation, $s_\phi$ in the GL region is proportional to the fluctuation entropy (i.e., the mean-squared pair-field) even in the quantum case, and the mean field 
result $\phi_0^2 \beta (1-T/T_{c0})/(16 \pi^2 \lambda^2(0))$ is expected to be recovered deep in the vortex liquid regime if the fluctuation has calmed down there. 

In the following sections, the above expressions of $\sigma_{\rm fl}$, $\sigma_{\rm vg}$, and $U_\phi$ are used together with eqs.(5), (8), (9), (10), (11), and (16) to examine experimental data of transport quantities, 
primarily, of the resistivity 
\begin{equation} 
\rho=(\sigma_n+\sigma_{\rm fl}+\sigma_{\rm vg})^{-1},
\label{21}
\end{equation} 
where the normal conductivity $\sigma_n(T)$ should be estimated through experimental $\rho(T)$ data in $T > T_{c0}$ by, as usual, neglecting its $H$-dependence and is implicitly assumed to include other fluctuation conductivity terms excluded from the GL description. The Nernst coefficient $N$, measured in refs.16 and 17, is defined as 
\begin{equation}
N=\frac{s_\phi \rho}{\phi_0}. 
\label{22}
\end{equation}
We note that, since $U_\phi$ is proportional to $-\phi_0 M$ in the GL region, the magnetization $M$ may be chosen in place of $U_\phi$ as a quantity to be compared with $\rho$. Even if taking account of electronic details of materials of interest, $\lambda(0)$, $T_0$, $H_0$, and $T_{c0}$ are independent SC material parameters in an ordinary microscopic description of superconductors. They appear in the coefficients of the GL action such as $\mu_0$ and $\gamma_0$. On the other hand, $\sigma_n(T)$ is highly sensitive to the sample purity and can be seen as being independent of the four SC parameters mentioned above. In order to mimic a vanishing behavior of $\rho(T)$ far below $T_{c0}$, a VG term $\sigma_{\rm vg}$, i.e., a (sample-dependent) vortex pinning effect on the conductivity, needs to be incorporated. If the 3D form, eq. (17), is used, its prefactor $c_p$ and a form of transition line $t_{\rm vg}(H)=T_{\rm vg}(H)/T_{c0}$ are chosen to optimize the fittings. In general, the vortex pinning in real systems may occur due to crystal defects other than a microscopic impurity affecting $\sigma_n$, and further, the transition line $t_{\rm vg}(H)$ may not be described precisely in the LLL approximation assumed so far, once recalling a strong $H$-dependence of the vortex elastic moduli and the presence in real samples of a small amount of other pinning sites. That is, the vortex pinning should be regarded as being independent of $\sigma_n$ and is not necessarily described in terms of a single extrinsic parameter. If the 2D form (18) is a more appropriate $\sigma_{\rm vg}$, the pinning strength $b_p$ is the only extrinsic (sample-specific) parameter at {\it nonzero} temperatures. 

\section{Cuprate Superconductors} 
In this section, the theoretical expressions in $\S 2$ will be applied to experimental data of superconducting cuprates. Since one purpose of examining resistivity data of cuprates is to correctly understand the doping dependences of fluctuation effects and of material parameters of cuprates, we will primarily examine resistivity data of LSCO of which an extensive doping dependence can be seen in the literature \cite{HS}. As shown in ref.20 where the expressions obtained in $\S 2$ were applied, data of other quantity measured consistently are also 
needed, 
together with resistivity data, to correctly estimate material parameters in cuprates with small condensation energy. In underdoped cuprates, such a set of data measured consistently are not known except the LSCO data in refs.16, 17, and 39. Below, we will proceed further the analysis to the overdope side and comment on other cuprate materials on the basis of available data \cite{Capan2,Wang1}. 

Let us start with incorporating microscopic ingredients into the GL description. Since experimental data in several teslas are primarily examined for materials with a much lower $H_{c2}^*(0)$ than the optimally-doped YBCO, we need a reasonable functional form of $H_{c2}^*(T)$ which may not be approximated in several teslas as the ordinary linear straight line. Further, the time scales $\gamma_n$ need to be calculated consistently with this $H_{c2}^*(T)$. To this end, we invoke the ordinary clean limit \cite{Lee} in order to describe $\mu_n$ and $\gamma_n$ consistently (see $\S 2$ on their definition). For simplicity, let us assume, as in the weak-coupling $s$-wave pairing case, a circular Fermi surface. Then, they are given by
\begin{eqnarray}
\gamma_n= {\beta \over {2 \pi}} \int_0^\infty ds {s \over {{\rm sinh}(s)}} {\rm L}_n(u_c^2 s^2) \, e^{-(u_c s)^2/2} \;, 
\\ \nonumber
\mu_n={\rm ln}\biggl({T \over {T_0}}\biggr) + \int_0^\infty ds \, {{1- {\rm L}_n(u_c^2s^2) \, e^{-(u_c s)^2/2}} \over {{\rm sinh}(s)}},
\label{23}
\end{eqnarray}
respectively, where $u_c=T_0 \sqrt{H/(2 H_0 e^\gamma)}/T$, ${\rm L}_n(x)$ is the $n$-th order Laguerre polynomial, and $\gamma=0.5771$ is the Euler constant which should not be confused with the time scales $\gamma_n$ \cite{com4,com3}.  
It is valuable to comment on the fact that $\gamma_n$ with any odd integer $n$ approaches zero in $T \to 0$. Since an equilibrium vortex solid state with no Pauli limiting effect is represented by the LLs with even $n$, a dissipative vortex flow motion is created by other odd LLs \cite{RI95} so that the vortex flow conductivity is proportional to $\gamma_n$ with an odd $n$. Hence, this result suggestive of dissipation-free vortex flow at $T=0$ in clean limit may be rather expected and significantly enhances the quantum {\it resistive} behaviors in cuprates and organics at {\it finite} temperatures. 

As noted in $\S 2$, the apparent upper critical field line $H_{c2}^*(T)$ approaching $T_{c0}$ in low $H$ limit (see Fig.3) is determined by $\mu_0+\Delta \Sigma_h=0$. When $\Delta \Sigma_h={\rm ln}(T_0/T_{c0})$, we obtain 
\begin{equation}
H_{c2}^*(T) = H_0 \biggl({{T_{c0}} \over {T_0}} \biggr)^2 \Phi(t),
\label{24}
\end{equation}
from eq.(23), while $H_0(T)$ is given by $H_0 \Phi(T/T_0)$, where the 
function $\Phi(x)$ satisfies $\Phi(0)=1$ and $\Phi(1)=0$. 
In particular, a large enough value of the parameter $T_0/T_{c0}$ significantly affects fluctuation phenomena in nonzero fields. 

Note that, when $T_{c0}$ rather than $T_0$ is chosen as a temperature parameter scaling $T$, $H_0$ is replaced by $H_{c2}^*(0) = H_0 (T_{c0}/T_0)^2$. By combining this with eqs.(4) and (11), we find the property 
\begin{equation}
\sigma_{\rm fl}(T_0/T_{c0}, H_0, \lambda(0)) 
= \sigma_{\rm fl}( 1, H_{c2}^*(0), T_{c0} \lambda(0)/T_0)  
\end{equation}
to be valid in the high $H$ approximation. 
Namely, if the SC pseudogap region with the width $T_0-T_{c0}$ is neglected, $\xi_0 = \sqrt{\phi_0/(2 \pi H_0)}$ is overestimated, while $\lambda(0)$ is underestimated. A discussion based on this fact will be given in $\S 6$. 

As the first example of comparisons with data, the case of optimally-doped YBCO shown in Fig.1 (a) will be explained. In this case, the $\rho(T)$ curves in the tesla range show the fan-shaped broadening in the thermally-induced vortex liquid region below $T_{c0}$, and the thermal SC fluctuation reduces $\rho$-values near $T_{c2}^*(H)$ compared with the mean field result of resistivity. Figure 1(a) includes 12 (T) data in ref.22 and $\rho(T)$ curves computed in terms of the material parameters given in Table I, while the curves in Fig.1 (b) result from the same set of material parameters, except $\lambda(0)=0.35$ ($\mu$m), as in Fig.1 (a). Further, the normal resistivity $\sigma_n^{-1} = 0.135 T/T_{c0}$(m$\Omega$.cm) and the 3D-like $\sigma_{\rm vg}$, eq.(17), with $c_p=1.3 \times 10^{-4}$ and $t_{\rm vg}=1-h-1.2 h^{2/3}$, where $h=H/H_{c2}^*(0) \ll 1$, were assumed. Since, in several teslas, the $h$-values of optimally-doped YBCO are much lower than those in most of underdoped and overdoped materials including LSCOs to be discussed below, an additional contribution \cite{RI1} composed only of {\it thermal} higher LL modes to $\sigma_f$ was also included in obtaining the solid curves together with the quantum contribution accommodated in eq.(15). In Fig.1 (a), however, a change of resistivity value brought by the addition of higher LL contributions is within several percents in magnitude, and the quantum contribution was quite negligible. 
The positions of $H_{c2}^*(T)$ for each resistivity curve are denoted by filled circles both in Fig.1 (b) and the figures appearing hereafter. 
Further, we note that, although a small but nonvanishing $T_0/T_{c0} - 1$ was taken into account in Fig.1 favoring an agreement with the consistent data of $U_\phi$ (see Fig.4), the obtained values of $\lambda(0)$ and $\xi_0$ remained unchanged compared with those in previous fittings \cite{RI1} where the quantum contribution was neglected from the outset. 
\begin{figure}[t]
\scalebox{0.4}[0.4]{\includegraphics{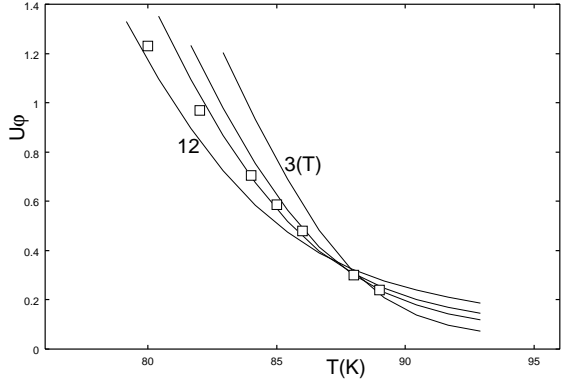}}
\caption{$U_\phi$-curves (solid curves) at $3$, $6$, $8$, and $12$ (T) obtained consistently with those in Fig.1 (a) and 12 (T) 
data (open squares) \cite{Huebner}. In this and other figures of $U_\phi(T)$, $U_\phi$ is represented in unit of $10^{-12}$(J/m).}
\label{} \end{figure}

We note that $T_0$ we define is the bare mean field SC transition point defined {\it prior to} including a coupling with possible non SC orderings competing with the SC ordering. We expect two types of origins of the SC pseudogap region $T_0-T_{c0}$ in cuprates. First, inclusion of a fluctuation of a competing non-SC ordering, such as a spin or charge ordering, coupling to the SC order parameter can lead to a reduction of the mean field (MF) SC transition point. The resulting MF transition temperature in $H=0$ will be called $T_c^{\rm MF}$ hereafter. This can be modelled by adding the term  
\begin{equation}
{\cal S}_x= u_x \int_\tau \int d^2r |\phi_{ns}|^2 |\psi|^2
\end{equation}
in GL action, where $\phi_{ns}$ denotes a non-SC order parameter fluctuation, and only a fluctuation competitive (or repulsive) to $\psi$ is assumed here through the condition $u_x > 0$. Since $|\phi_{ns}|^2$ is replaced by the averaged value $<|\phi_{ns}|^2>$ in constructing an effective action on $\psi$, $T_c^{\rm MF}$ will be expressed as $T_0 \exp(-u_x <|\phi_{ns}|^2>)$ ($< T_0$), and consequently, we has only to replace ${\rm ln}(T/T_0)$ in $\mu_0$ by ${\rm ln}(T/T_c^{\rm MF})$, where $<|\phi_{ns}|^2>$ was assumed to be $H$-independent. The second origin of $T_0$-shift is nothing but $\Delta \Sigma_h$ in the LLL mass renormalization outlined in $\S 2$, where it was assumed to arise entirely from the SC fluctuation in higher LLs. An additional contribution to $\Delta \Sigma_h$ also arises from ${\cal S}_x$ and similar higher order coupling terms between $\psi$-fluctuations and $\phi_{ns}$. As in $\S 2$, by assuming this contribution to $\Delta \Sigma_h$ of such a $\phi_{ns}$-fluctuation to be also $H$-independent, $\Delta \Sigma_h$ can be identified with ${\rm ln}(T_c^{\rm MF}/T_{c0})$. That is, $T_c^{\rm MF}$ does not appear in $\mu_0+\Delta \Sigma_h$. 
\begin{table}
\begin{tabular}{cccc}
&Fig.1(a), 4 &Fig.5 &Fig.8 \\ 
\hline
$\lambda$(0) ($\mu$m) & 0.11 & 0.43 & 1.9 \\
$H_0$ (T) &272 &235 &330 \\
$T_0$ (K) &120 &96 &90 \\
$T_{c0}$ (K) &92 &32 &15 \\
$d$ (nm) &1.5 &1.5 &1.5 \\
\hline
\end{tabular}
\caption{Material parameter values used in Figs.1, 4, 5, and 8. Parameters related to the pinning effect and $\sigma_n$ are given in the text.}
\end{table}
\begin{figure}[t]
\scalebox{0.42}[0.42]{\includegraphics{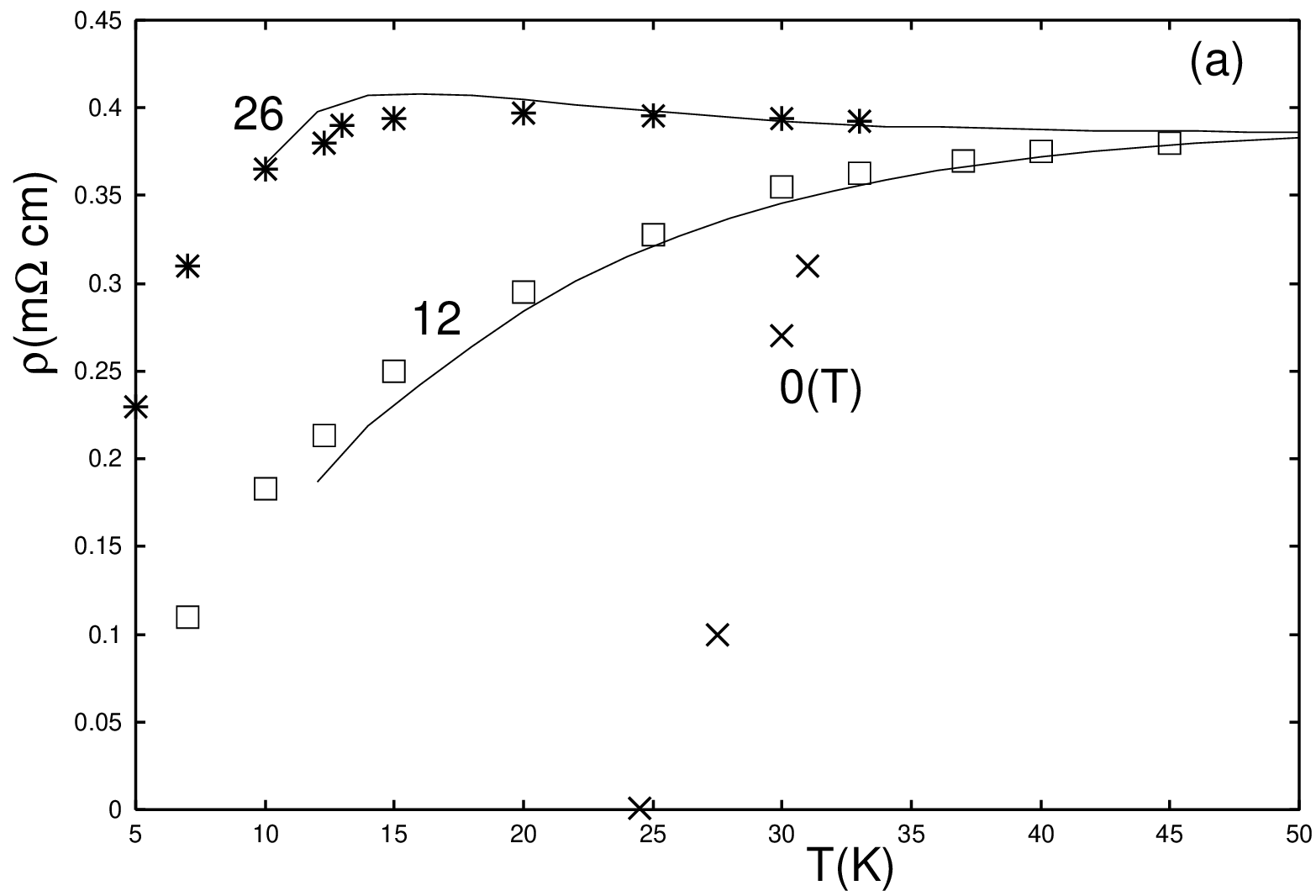}}
\scalebox{0.4}[0.4]{\includegraphics{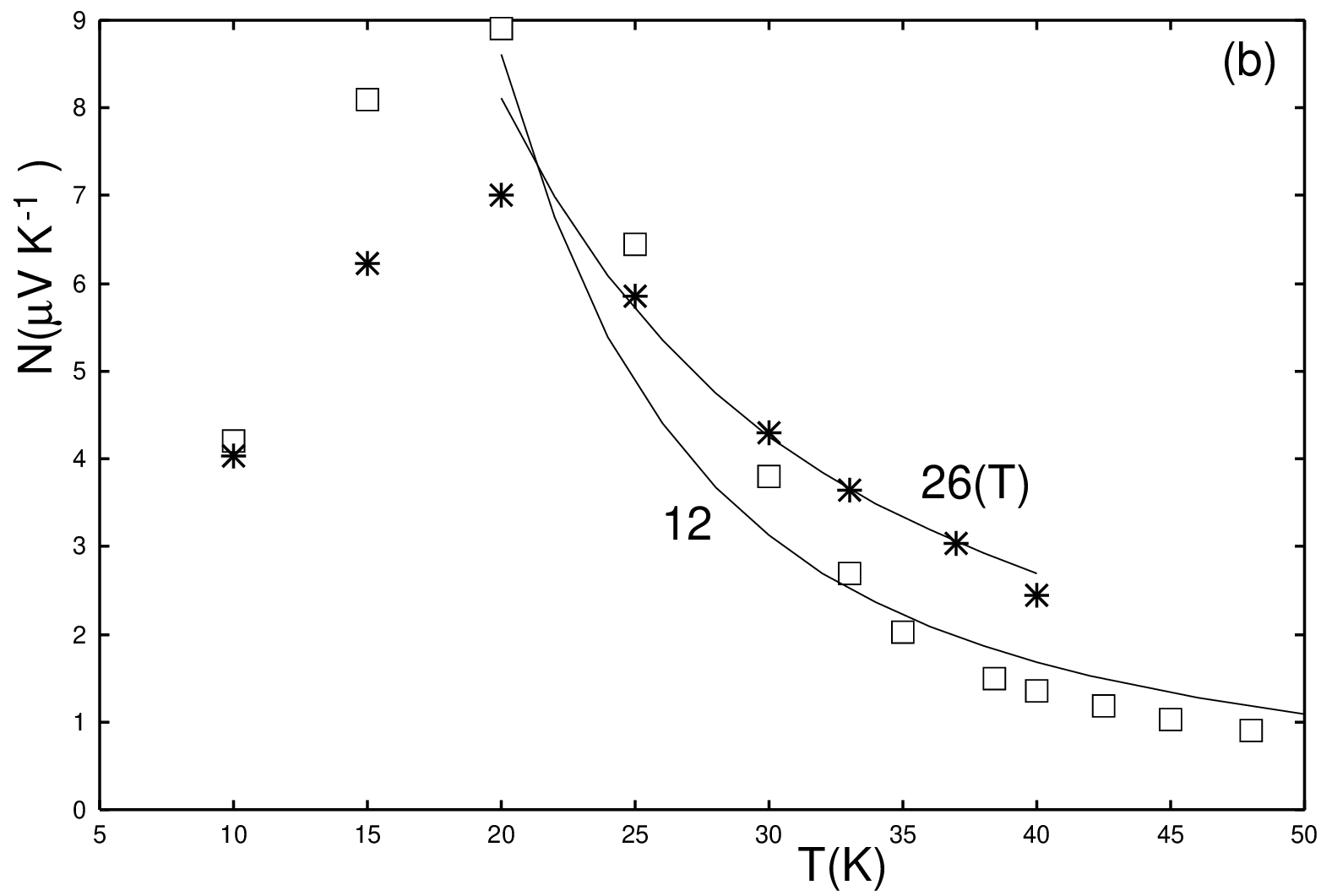}}
\caption{Fittings to LSCO $x=0.08$ data at $12$ and $26$ (T) of (a) $\rho$ and (b) $N$ in Ref.16.}
\label{} \end{figure}

The assumption of $H$-independence of the $\phi_{ns}$-fluctuation can be justified by comparing with experimental observations as follows. First, according to NMR measurements \cite{NW} in the pseudogap regime, the antiferromagnetic (AF) fluctuation, which is competitive with the SC fluctuation and should be dominant in the NMR signal in the nodal directions, is suggested to be insensitive to $H$ at least in $T > T_{c0}$. On the other hand, a local AF ordering was shown to be enhanced with increasing $H$ primarily below the irreversibility line where the ohmic resistivity is absent \cite{MIT}. This $H$-dependence of AF ordering is essentially linear in $H$, reflecting the number of vortices, and possibly, will be a consequence of the spatial variation of $|\psi|$ near the vortex cores \cite{Kivelson,Sachdev}. However, in the high $H$ approach of GL theory, the spatial variation of $|\psi|$ is reflected only in $\beta_A - 1$ which is negligible \cite{RI3} in discussing thermodynamics and transport phenomena above the irreversibility line (i.e., in the vortex liquid regime). Therefore, we believe that the assumption of $H$-independence of $\phi_{ns}$-fluctuation is valid in the vortex liquid regime. Even if the $\phi_{ns}$-fluctuation induces an $H$-dependence in the effective action on the $\psi$-fluctuation, both fluctuations are competitive with each other, and hence, a SC coherence length defined from a $H$-dependence in the effective action would be enhanced by the presence of $\phi_{ns}$-fluctuations. As is repeatedly seen hereafter, however, $\xi_0$-values resulting from the fittings to data of underdoped cuprates already become short enough. For this reason, the extra $H$-dependence due to the $\phi_{ns}$-fluctuations is likely to be quantitatively negligible, and hence, we assume the contributions of $\phi_{ns}$-fluctuation to have been fully accommodated as ${\rm ln}(T_0/T_c^{\rm MF})$. Although, as a result of this, $T_c^{\rm MF}$ does {\it not} appear in the r.h.s. of eqs.(4) and (9), the mean field transition field in the presence of $\phi_{ns}$-fluctuations will be given, consistently with eq.(24), by $H_0 (T_c^{\rm MF}/T_0)^2 \Phi(T/T_c^{\rm MF})$ which may decrease with underdoping. Note, nevertheless, that the genuine $\xi_0$ is defined through the microscopic field scale $H_0$ {\it irrespective of} $\phi_{ns}$-fluctuations. 
\begin{figure}[t]
\scalebox{0.4}[0.4]{\includegraphics{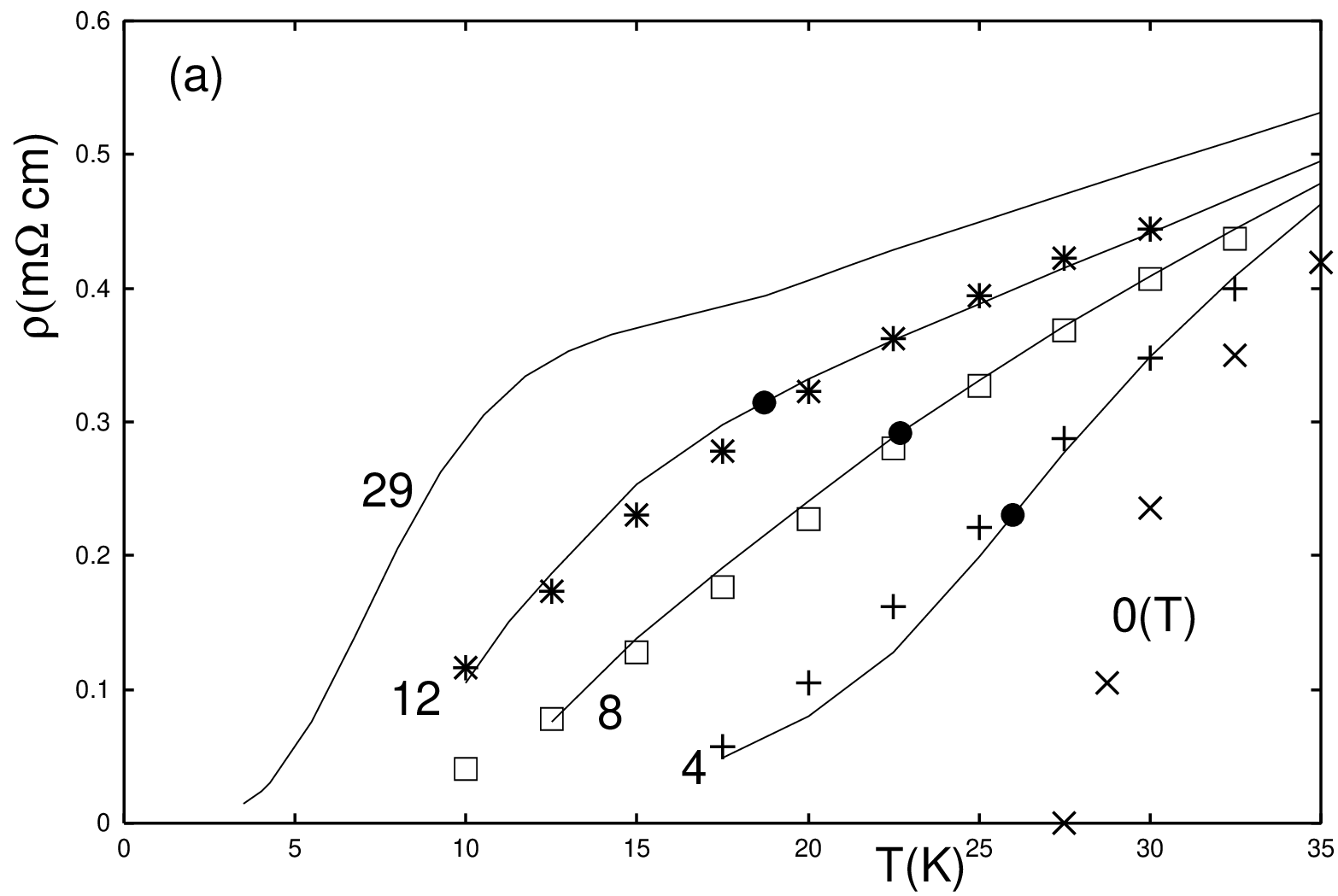}}
\scalebox{0.4}[0.4]{\includegraphics{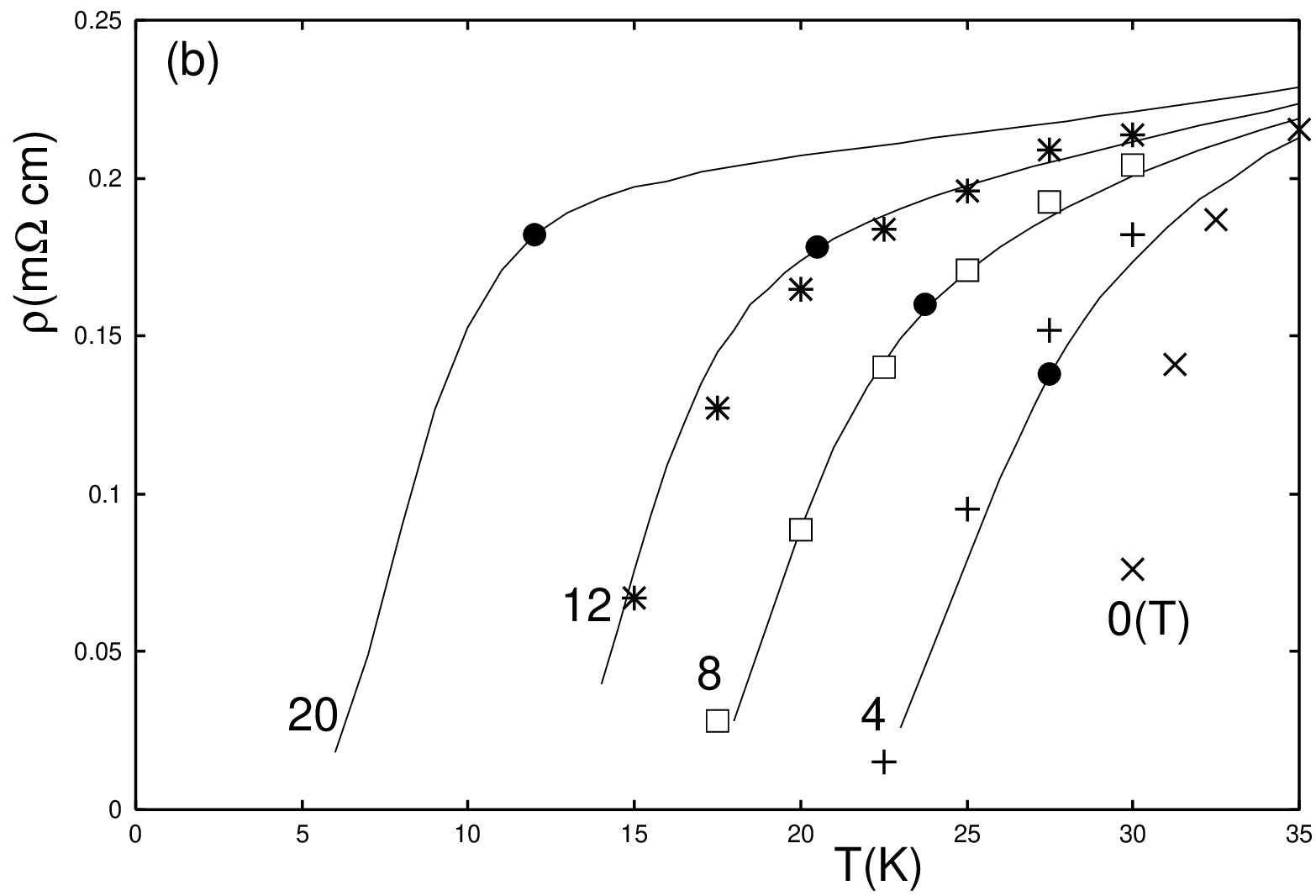}}
\scalebox{0.4}[0.4]{\includegraphics{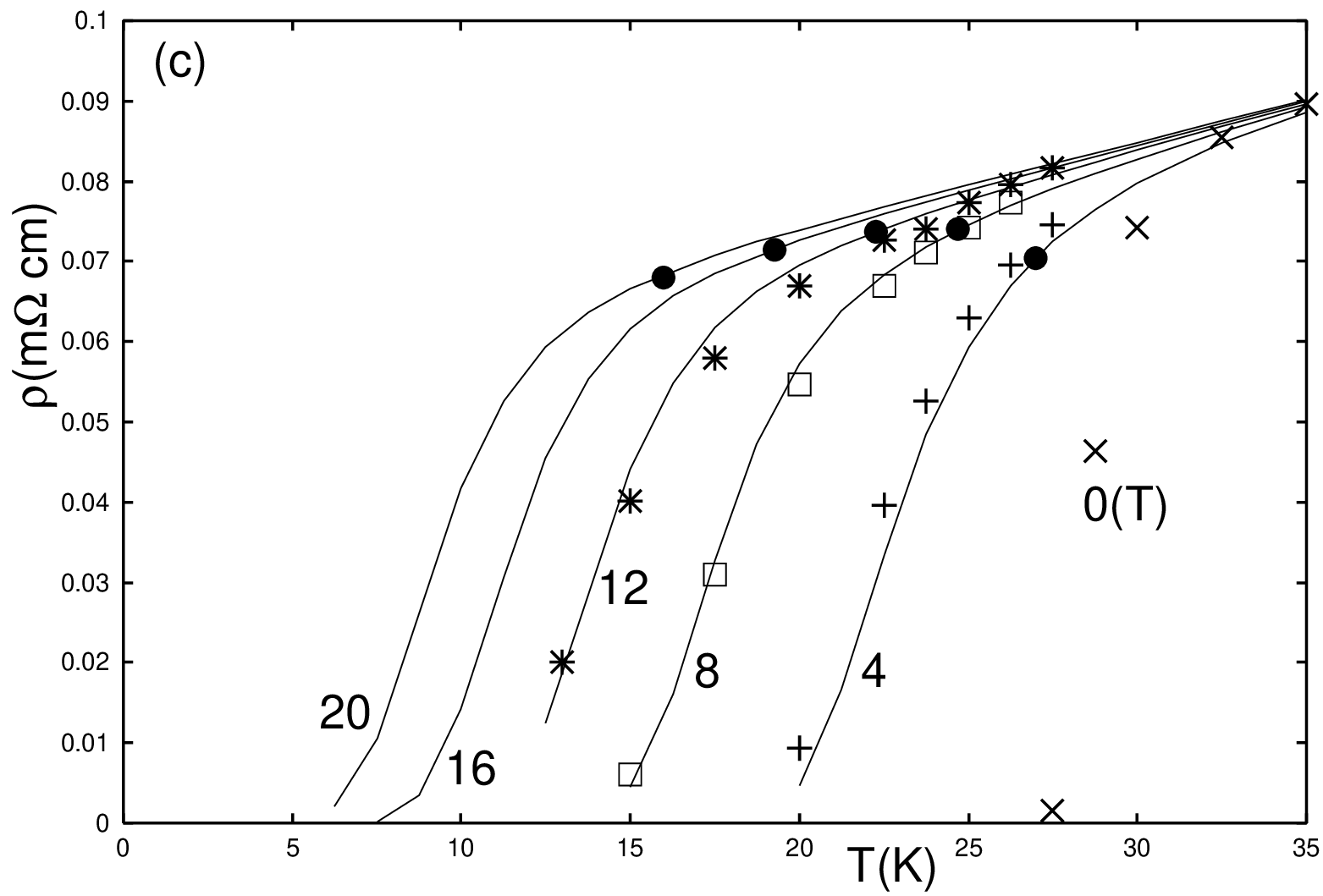}}
\caption{Resistivity data (symbols) \cite{HS} of LSCO in the same temperature range at three doping levels (a) $x=0.08$, (b) $0.15$, and (c) $0.2$ and corresponding theoretical results (solid curves). The darked circle on each $\rho(T)$ curve denotes $T_{c2}^*(H)$ in each $H$.}
\label{} \end{figure}
\begin{figure}[t]
\scalebox{0.35}[0.35]{\includegraphics{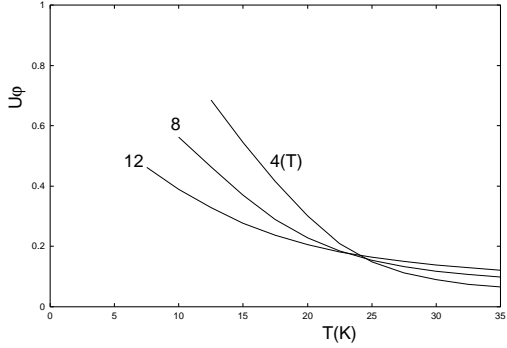}}
\caption{$U_\phi$-curves corresponding to $\rho$-curves in Fig.6 (c). }
\label{} \end{figure}

Based on the inclusion, explained above, of competing $\phi_{ns}$-fluctuations,  the fitting results to LSCO$x$ data with $x=0.06$ and $0.08$ reported in ref.20 will be commented on here. Hereafter, in considering the doping dependence of cuprates, the spacing $d$ between the SC layers will be fixed. In ref.20, the formulas given in $\S 2$ were applied by neglecting the difference between $T_0$ and $T_c^{\rm MF}$. Regarding $\sigma_{\rm vg}$, eq.(17) with $c_p= 10^{-2}$ was used for $x=0.08$ case, and eq.(18) with $b_p= (2 \pi \xi_0^2 d) 0.023$ was applied to $x=0.06$ case . According to the results in ref.20, $T_0(x)$ became insensitive to $x$ in $x < 0.1$ in contrast to $T_\nu$, defined in ref.26 as the onset of SC Nernst signal, which {\it decreases} with decreasing $x$ ($< 0.1$). Here, in relation to Fig.6 (a) below, the fitting results to the $x=0.08$ data given as Fig.2 in ref.20 will be shown again in Fig.5, where $(\sigma_n)^{-1}=0.02905 \, {\rm ln}(1.6 \times 10^6 T_{c0}/T)$ (m$\Omega$.cm) and the parameters listed in Table I were used. Regarding the $x=0.06$ case, it has been noticed recently that, even if assuming $T_0$ to increase with decreasing $x$ just like $T_\nu$ in Bi-compounds \cite{Wang1}, the data can be quantitatively explained. For instance, the parameter values $\lambda(0)=2.3$($\mu$m), $T_0=100$(K), $T_{c0}=13$(K), $H_0=493$(T), and $(\sigma_n)^{-1}=0.0994 {\rm ln}(186.3 T_{c0}/T)$ (m$\Omega$.cm) were used for the $x=0.06$ case, and the obtained curves of $\rho$ and $N$ were almost the same as those in Fig.1 of ref.20 where a smaller $T_0$ value, $96$(K), was assumed. Contrary to this, no choice of $T_0$ decreasing with underdoping has resulted in a consistency with the Nernst data \cite{Capan1} particularly in the $x=0.06$ case. What corresponds to $T_\nu$ in low $H$ limit is the mean field transition point $T_c^{\rm MF}$. More accurately, because the Nernst signal due to {\it Gaussian} SC fluctuation is usually nonzero above the microscopic transition point, $T_c^{\rm MF}$ may lie slightly below $T_\nu$. Hence, $T_\nu$ in $x < 0.1$ of LSCO, showing the unexpected doping dependence, will correspond to $T_c^{\rm MF}$ affected by competing non-SC fluctuations becoming stronger with underdoping. That is, we expect $T_0$ in LSCO to, contrary to $T_c^{\rm MF}$, 
increase with underdoping like in the Bi-compounds \cite{Wang1}. 
\begin{table}
\begin{tabular}{cccc}
$x$ &$0.08$ &$0.15$ &$0.2$\\
\hline
$\lambda$(0) ($\mu$m)& 0.57 & 0.46 & 0.417 \\
$H_0$ (T)& 245 & 190 & 55.1 \\
$T_0$ (K)& 100 & 86.5 & 40.5 \\
$T_{c0}$ (K)& 30 & 31 & 30 \\
$d$ (nm)& 1.5 & 1.5 & 1.5 \\
$\sigma_n^{-1}$ (m$\Omega$ cm)& 0.245 $\!+ \!$ 0.27$t$ & 0.19 $\! + \!$ 0.0426$t$ & 0.055 $\! + \!$ 0.0312$t$ \\
$c_p$ & 0.04 & 0.01 & 0.033 \\
$\alpha_{\rm vg}$ & 0.5 & 0.67 & 0.67 \\
$\beta_{\rm vg}$ & 3.4 & 0.8 & 2.26 \\
$H_c$(0) (T)& 0.352 & 0.385 & 0.23 \\
\end{tabular}
\caption{\label{tab:table2}Parameter values used in computation of resistivity curves in each figure of Fig.6. The parameter $t$ denotes $T/T_{c0}$.}
\end{table}

Next, in relation to Fig.5, the doping dependence of resistivity curves will be examined in order to corroborate that doping dependences of material parameters predicted in ref.20 are valid over a wider doping range including the overdoped side. Previously, a systematic study of resistive behaviors at various doping levels of LSCO was reported \cite{HS}, and the data in ref.8 will be used here. We focus on the temperature range $10 < T$(K) $< 35$ and 
approximate each $[\sigma_n(T)]^{-1}$ curve there via a $T$-linear curve. Among various data in ref.8, we have examined three doping levels, $x=0.08$ 
(underdoped case), $x=0.15$ (nearly optimal case), and $x=0.2$ (slightly 
overdoped case). Fluctuation effects in strongly overdoped materials ($x \geq 0.24$) are expected to be quite weak, and those data will not be considered here. First of all, the $x=0.08$ data have been fitted by assuming a $T_0/T_{c0}$ value similar to that in Fig.5. Next, the parameter values for $x=0.2$ case were chosen by favoring a semiquantitative agreement with Nernst data in ref.17 (see Fig.1 there). Then, in $x=0.15$ case, a $T_0/T_{c0}$ value intermediate between the $0.08$ and $0.2$ cases was assumed. However, according to eq.(25), the key parameter in comparing with $\rho$ data is the product $\lambda(0) \xi_0$, that is, $H_c(0) = (\lambda(0))^{-1} \sqrt{\phi_0 H_0/4 \pi}$ (see eq.(2)) which is independent of $T_0$ in GL theory. The obtained curves are shown together with the data \cite{HS} in Fig.6, and the values of material parameters used for fittings are listed in Table.II. The parameters $\alpha_{\rm vg}$ and $\beta_{\rm vg}$ are included in the assumed form of VG transition curve $t_{\rm vg}(h)=T_{\rm vg}(h)/T_{c0} = (1-h)/(1+\beta_{\rm vg} h^{\alpha_{\rm vg}})$, where $h=H/H_{c2}^*(0)$. 
This expression of $t_{\rm vg}(h)$ is based upon the VG transition line in the mean field approximation derived within the LLL \cite{RI3}. The factor $1-h$ arises from the mean-squared amplitude of the pair-field, while the exponent $\alpha_{\rm vg}$ depends upon the dimensionality of SC fluctuation and takes $0.5$ and $0.67$ in its 2D and 3D limits, respectively. This $\alpha_{\rm vg}$ is a direct measure of the sample anisotropy in the present approach applicable primarily to 2D-like materials (see the second paragraph of $\S 2$). The $\alpha_{\rm vg}$ values shown in Table II are compatible with the well-known fact that the hole-doped cuprate materials are more 2D-like with underdoping. By contrast, $\beta_{\rm vg}$ is sample-specific and also depends on both of the fluctuation strength and the pinning strength. Furthermore, there are at least two ingredients affecting the form of $t_{\rm vg}(H)$ in realistic cases. One is an ingredient independent of pinning effects and induces a deviation of $t_{\rm vg}$-form from its LLL expression. Physically, a detail of the vortex elasticity, such as the $H$-dependence of the shear modulus of a vortex lattice defined locally, will affect the form of $t_{\rm vg}(H)$, and hence, it is not surprising that such a difference of the local shear modulus in $h < 1$ from its LLL expression 
affects 
the $t_{\rm vg}(H)$-expression. Secondly, a small amount of line-like (or plane-like) pinning disorders in real systems also affect the transition line. 
For instance, in real systems including both point-like and line-like pinning disorders, the functional form of $t_{\rm vg}(h)$ depends even on the $h$-values \cite{RI3}. Thus, in fitting to resistivity data, it may be rather necessary to include a possible deviation of the $t_{\rm vg}(H)$-form from the expression in LLL and in the purely point disorder case. To reduce the fitting parameters as far as possible, we have assumed that such involved ingredients will be incorporated by taking $\beta_{\rm vg}$ as a sample-specific fitting parameter independent of other material parameters. 
Nevertheless, it should be stressed that the details of the extrinsic parameters $c_p$ and $t_{\rm vg}(h)$ should not be important for our purpose in this section of clarifying the doping dependence of {\it intrinsic} material parameters which are determined, roughly speaking, through the (pinning-independent) upper-half of the resistivity curves. 

In lower fields in $x=0.08$, the resistivity curves have the fan-shaped 
broadening suggestive of a dominance of thermal fluctuation over the quantum contribution. One might wonder if the result in Fig.6 (a) that the thermal (fan-shaped) behavior in lower fields is more evident with underdoping is consistent with the result \cite{RI7} on {\it strongly} underdoped ($x < 0.1$) cases \cite{Mac2,Capan1} where the thermal behavior was lost with underdoping in $x < 0.1$. This apparently conflicting result is resolved as follows. In the case of underdoped cuprates in tesla range, a strong SC fluctuation near $T_0$ is weakened to some extent upon cooling down to $T_{c0}$ and behaves, much below $T_0$, like that in effectively higher fields. Roughly speaking, such a reduction of fluctuation arises from an increase, upon cooling in $T_{c0} < T 
< T_0$, of the coefficient of the gradient term in the $H=0$ GL-expression. That is, the enhanced quantum fluctuation due 
to $\lambda(0)$ increasing with underdoping is partly compensated by the existence of a large SC pseudogap region which, in turn, results in weakening the fluctuation near and below $T_{c0}$. Hence, due to a $T$-dependence of an electronic origin, a long $\lambda(0)$ may not necessarily result in a quantum fluctuation-dominated behavior near $T_{c0}$ in the tesla 
range. Actually, in contrast to LSCO in $x > 0.1$, the quantum fluctuation effect on $\rho(T)$ seems to be {\it monotonously} enhanced with underdoping in any $H$ in the case of YBCO \cite{Mac2,Ando}. 

In Fig.6 (a), no $H_{c2}^*(T)$ position (dark circle) on the 29(T) curve was indicated. Actually, $H_{c2}^*(0)$ for Fig.6 (a) is close to 21(T), and hence, this 29 (T) curve is an example of the case in which the pinning-induced drop of resistivity occurs above $H_{c2}^*(T)$ as a consequence of a broad SC pseudogap region. It is not surprising because a VG transition can occur anywhere in $T < T_0(H)$, i.e., as far as the SC fluctuation is present. A similar feature will be discussed again in $\S 6$. 

In contrast, the resistivity curves in $x=0.2$ and $0.15$ cases always show a sharp drop. As suggested in ref.17, the high field $\rho$ curves in $x=0.2$ case show a sharp drop far below $H_{c2}^*(T)$ (dark circle) suggested from Nernst data, and hence the situation is likely to be similar to Fig.1 (b) in Introduction. Actually, the much weaker $H$-dependence of $\rho$ around $T_{c0}$ in $x=0.2$ compared with that in $x=0.15$ case is apparently inconsistent with an expected growth of $\xi_0$ accomanying the overdoping and rather reflects a fluctuation enhanced with overdoping. This is due to $\lambda(0) \xi_0$ increasing with overdoping starting from the optimal-doping, although the $H=0$ fluctuation is weakened with overdoping (see $\S 1$). 
The $U_\phi$-curves computed consistently with Fig.6 (c) are shown in Fig.7. The obtained $U_\phi$ values and $H$-dependences semiquantitatively agree with the data shown in Fig.1 of ref.17. Properties of the $U_\phi$ curves are typically thermal, and, at least below 12 (T), there are no remarkable quantum fluctuation effect on $U_\phi$. Regarding the $x=0.15$ case, we note that the VG transition position $t_{\rm vg}(h)$ is closer to $H_{c2}^*(T)$-line compared with those in $x=0.08$ and $0.2$ cases, and that, due to this, the sharp $\rho$-drop in $x=0.15$ case is rather similar to that in the mean field-like case with negligible vortex liquid region. This narrow vortex liquid regime suggested by the $x=0.15$ data may be realized if the effective pinning strength {\it relative to} the fluctuation strength is maximal near the optimal doping. 

One of important consequences arising from the fittings is that $[H_c(0)]^{-1} \propto \lambda(0) \xi_0$, i.e, the fluctuation strength in fixed $H$-values (see eq.(2)), is mimimal near the optimal doping and, just like $(T_{c0})^{-1}$, increases with {\it both} underdoping and overdoping from the optimal case. This is qualitatively consistent with the doping dependence of condensation energy density $[H_c(0)]^2/(4 \pi)$ estimated from the heat capacity data \cite{Loram}. It is easily understood by recalling the discussion on eq.(2) in $\S 1$ that the doping dependence of $\rho(T)$ curves mentioned in $\S 1$, including the fact that the fan-shaped resistive broadening is typically seen only near the optimal doping, is a reflection of this doping dependence of $H_c(0)$. Further, as Table II shows, the in-plane coherence length $\xi_0$ to be defined from $H_0$ monotonously decreases with underdoping over all doping ranges including $x < 0.1$ \cite{RI7}. This conclusion cannot be reached once the presence of the SC pseudogap region widening with underdoping is neglected \cite{Ando}. 

Examples of $\rho$ and $U_\phi$ curves in a case with very large strengths of {\it both} the SC fluctuation and pinning effect are shown in Fig.8. They have been computed by bearing very underdoped Bi-2201 data \cite{Capan2} in mind (see Fig.2 in ref.39). A quantitative comparison with the data will not be attempted here because the data were taken on a film sample with a broadening of $\rho(H=0)$ curve over 10 (K). This sample-specific broadening at $H=0$ should be also reflected in low $H$ curves of resistivity and make comparison of computed curves with the low $H$ data difficult. Nevertheless, we expect semiquantitative features of the Bi-2201 data except $\rho$-curves below 6(T) to be comparable with Fig.8. As well as in the LSCO case with $x=0.06$ \cite{RI7}, the 2D $\sigma_{\rm vg}$ expression, eq.(18), was used with $b_p=(2 \pi \xi_0^2 d) 0.15$. In Fig.8, $(\sigma_n)^{-1}=0.21 \, {\rm ln}(280 \, T_{c0}/T))$ (m$\Omega$.cm) was used together with the parameters shown in Table I. Qualitatively, the features are also similar to the LSCO data in $x \leq 0.06$ \cite{Karpinska,RI7}. First, the quantum SC fluctuation in Fig.8 is strong. In fact, $T_{c2}^*$ at 8(T) is close to 5(K) where the $\rho(T)$-curve is insulating. More notably, the resistivity curves above 5(K) suggest a 2D FSIT with $B_c^* \simeq  6$(T), like in Fig.1 of ref.20. However, the FSIT behavior in Fig.8 is more remarkable compared to the LSCO case with $x=0.06$ \cite{RI7}. This is a consequence of the larger value of pinning strength $b_p$. Namely, a large enough value of pinning strength is needed {\it together with} a strong quantum SC fluctuation to obtain a more remarkable FSIT behavior visible even near $T_{c0}$. On the other hand, the larger pinning effect enhances the VG transition field at low $T$. Actually, $H_{c2}^*(0)$ in Fig.8 is less 
than 
10(T) at which a rapid drop of $\rho$ at a finite $T$ still occurs. 
\begin{figure}[t]
\scalebox{0.4}[0.4]{\includegraphics{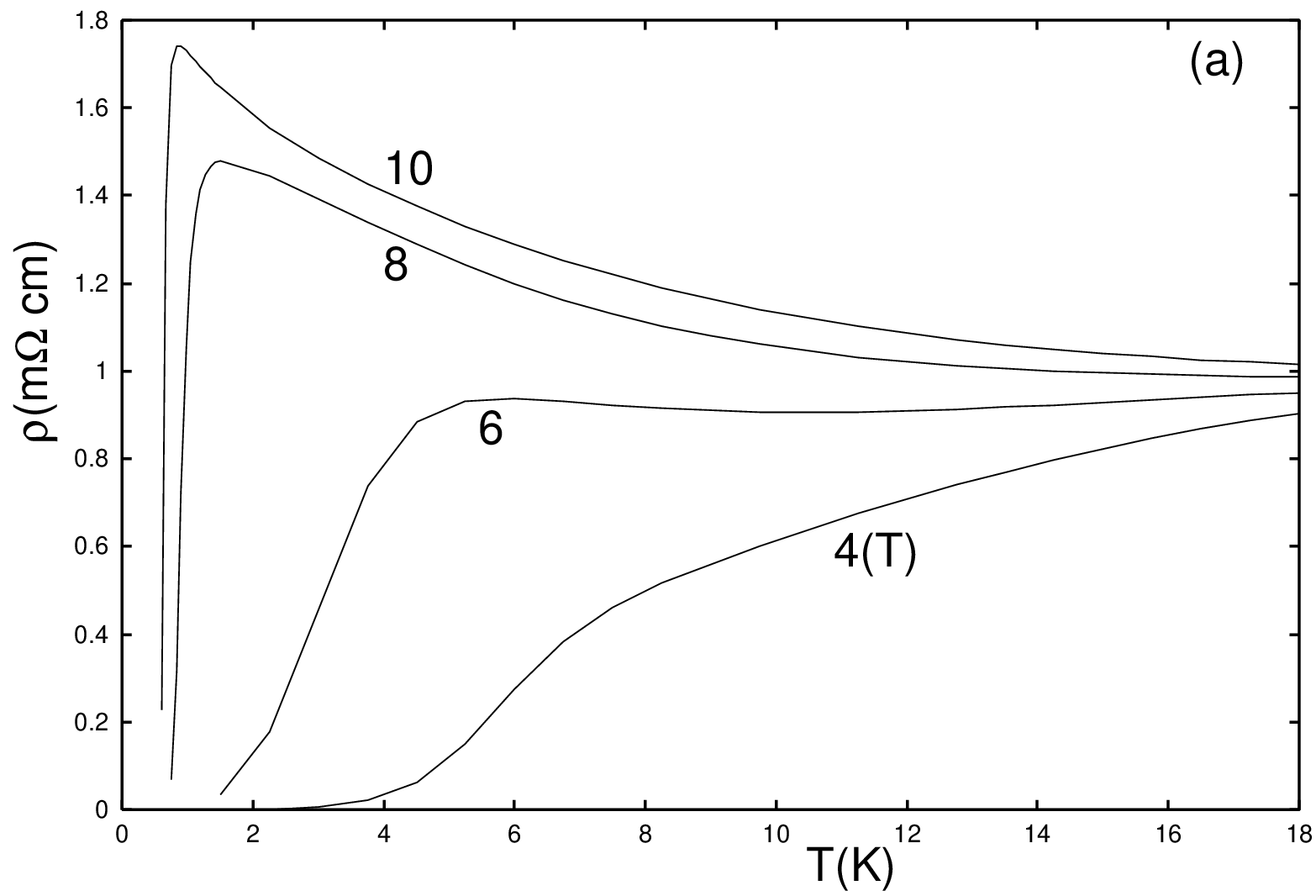}}
\scalebox{0.4}[0.4]{\includegraphics{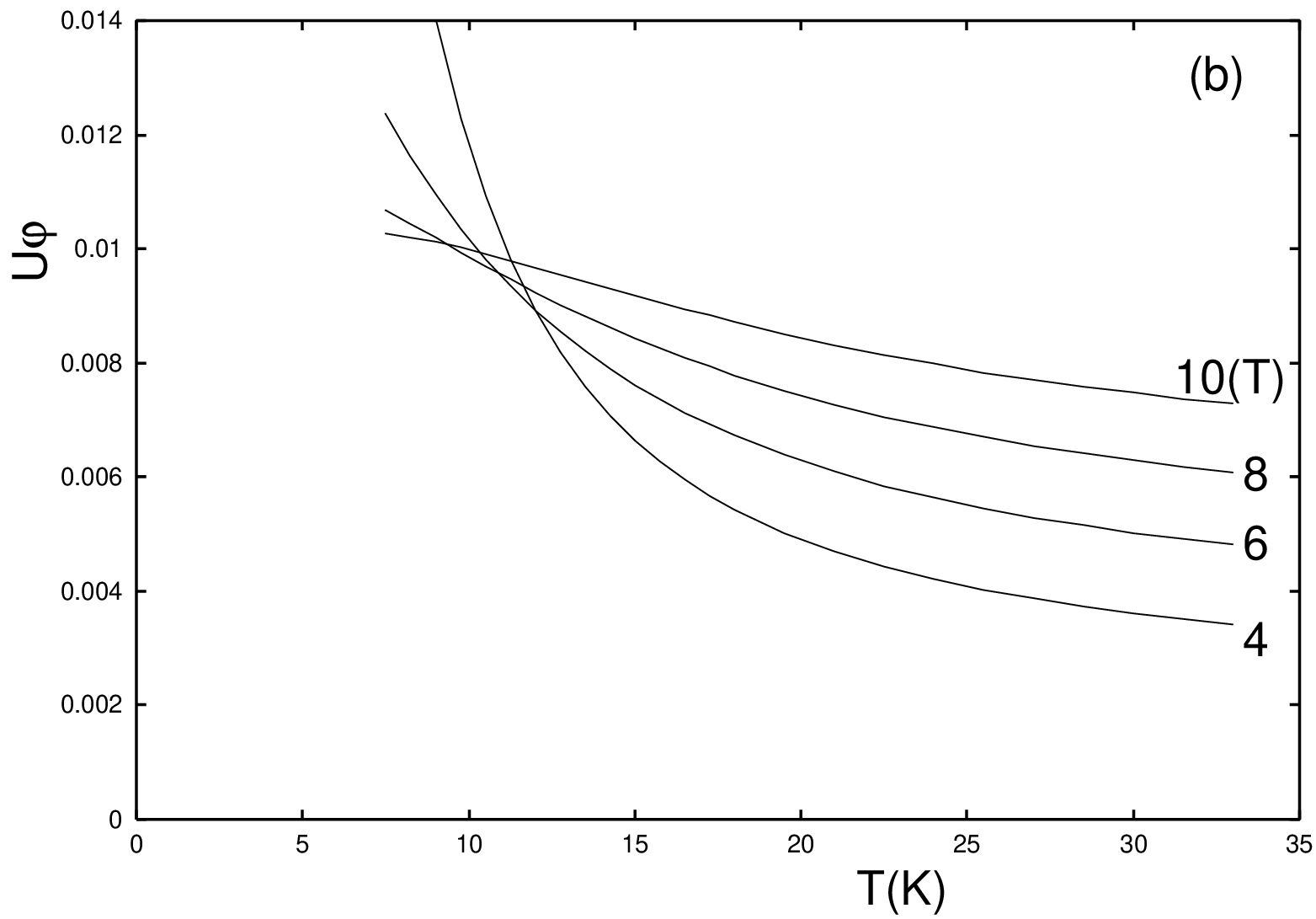}}
\caption{(a) Resistivity curves and (b) $U_\phi$-curves obtained by imagining underdoped Bi2201 data \cite{Capan2}. We note that $T^*_{c2}(H \leq 10({\rm T})) > 8$(K). }
\label{} \end{figure}

Regarding the transport energy, the $U_\phi$-values in Fig.8 (b) are quite low and should be compared with other figures of $U_\phi$ shown in this paper. As well as the LSCO case \cite{RI7}, the $U_\phi$ or $N$-values are two order of magnitude lower than a value expected by ignoring the SC pseudogap region (i.e., when $T_0=T_{c0}$) and are consistent with the data \cite{Capan2}. 

This comparison with underdoped Bi-2201 data corroborates the argument in ref.20 based on the LSCO data as follows. Since, in contrast to the LSCO case in $x < 0.1$, $T_\nu$ in Bi2201 seems to monotonously increase \cite{Wang1} with underdoping, one might expect underdoped Bi-data to behave in a qualitatively different way from the LSCO 
data. As seen above, however, the present theory in which $T_c^{\rm MF}$ corresponding to $T_{\nu}$ plays no essential roles explains the similarity \cite{Capan1,Capan2} in behaviors near and below $T_{c0}$ of both $\rho$ and $U_\phi$ data between Bi-2201 and LSCO. It implies that the examples of LSCO studied in ref.20 can be seen as generic behaviors of strongly underdoped cuprate 
materials below $T_{c0}$. 

In this paper, the case of underdoped YBCO is not examined in details because no comparable data of $\rho$ and $N$ at the same doping level in underdoped YBCO have been reported. Actually, the 3D nature of SC fluctuation should be incorporated in theoretical descriptions in contrast to other cuprates which are commonly much more 2D-like, and hence, the approach in $\S 2$ may not be directly applicable to a quantitative study of doping dependences of YBCO. Nevertheless, the following features are suggested from available $\rho$ and $N$ data: First, the fact that, in contrast to the LSCO $x=0.08$ data \cite{HS} showing the fan-shaped broadening below 8 (T), the $\rho$ data in underdoped YBCO \cite{Mac2,Gan,Ando} entirely show a sharp drop implies that the enhancement of quantum fluctuation always overcomes a reduction of fluctuation upon cooling arising from a broad SC pseudogap regime, and hence, a $T_0-T_{c0}$ value much smaller than that 
in LSCO 
is 
expected. This narrower SC pseudogap regime is presumably consistent with the Nernst coefficient \cite{Wang2} larger than in LSCO and Bi-2201. In any case, the doping dependences of $T_\nu$ and $T_0$ in YBCO, which we expect will be remarkably different from each other, should be 
clarified elsewhere. 

\section{Organic Superconductors}
Previously, the resistive behaviors in the vortex liquid regime of $\kappa$-(ET)$_2$ organic superconductors have been studied in parallel with those of cuprate materials. Typical data are seen in refs.18 and 19. Surprisingly, the resistivity curves in $\kappa$-(ET)$_2$Cu(NCS)$_2$ with a wider vortex liquid regime have shown a sharp drop near the irreversibility line in {\it all} fields shown there. In $\kappa$-(ET)$_2$Cu[N(CN)$_2$]Br with a narrower liquid regime, the resistivity $\rho(T)$ curves have shown a clear $H$-dependent crossover from the familiar fan-shaped broadening in lower $H$ into a sharp drop in higher $H$ near the irreversibility line lying much below $H_{c2}^*(T)$ estimated from the magnetization data. These features are much the same as those seen in cuprates \cite{HS,Mac2,Mac3} and are consequences of quantum SC fluctuation becoming more important as the fluctuation is stronger. Here, in addition to reproducing and discussing the fitting result,\cite{RI6} the low $T$ behaviors will be examined. Numerical results in this section are also based on the use of eqs.(23) and (24) within the framework of $\S 2$. 
\begin{figure}[t]
\scalebox{0.45}[0.45]{\includegraphics{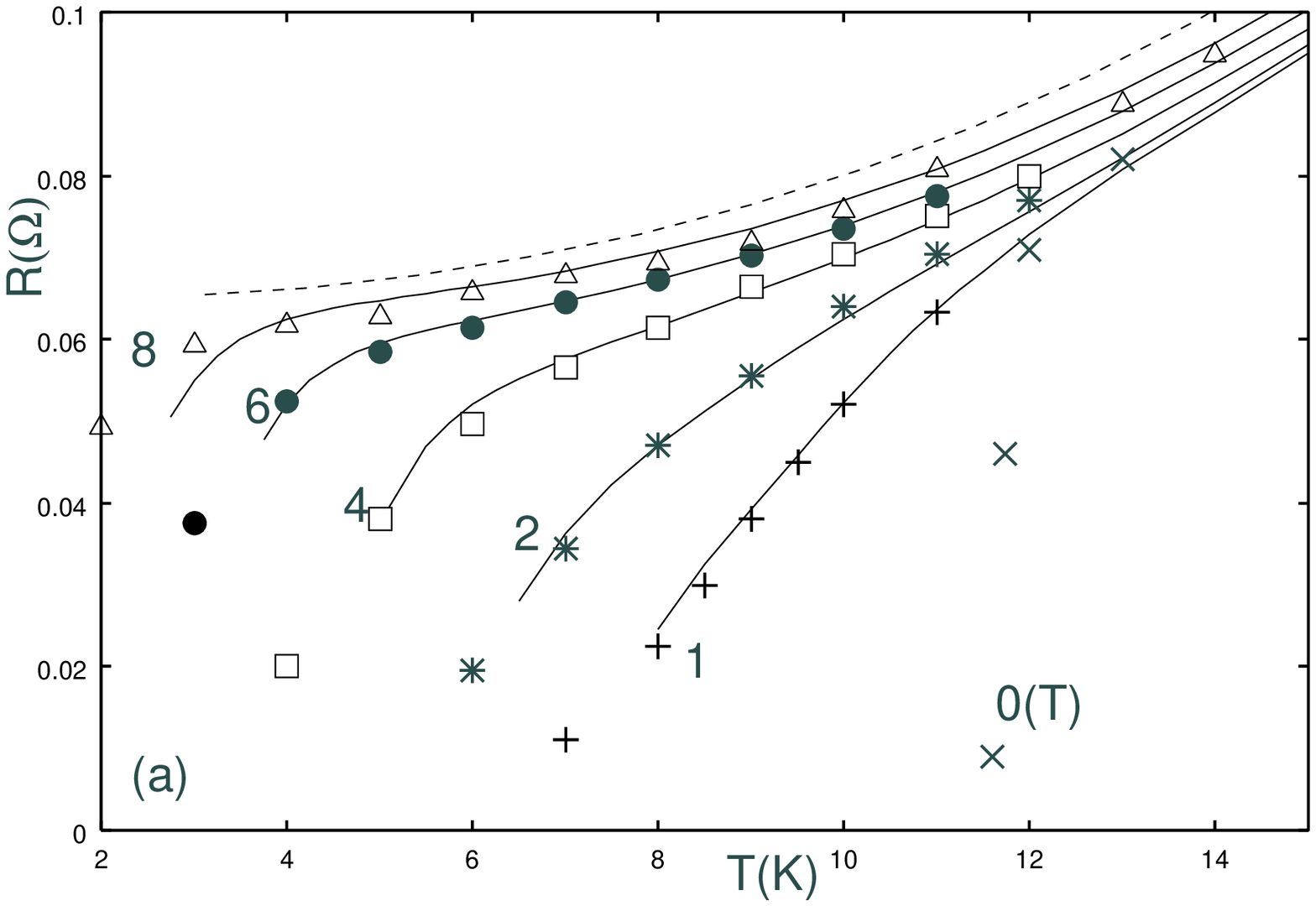}}
\scalebox{0.45}[0.45]{\includegraphics{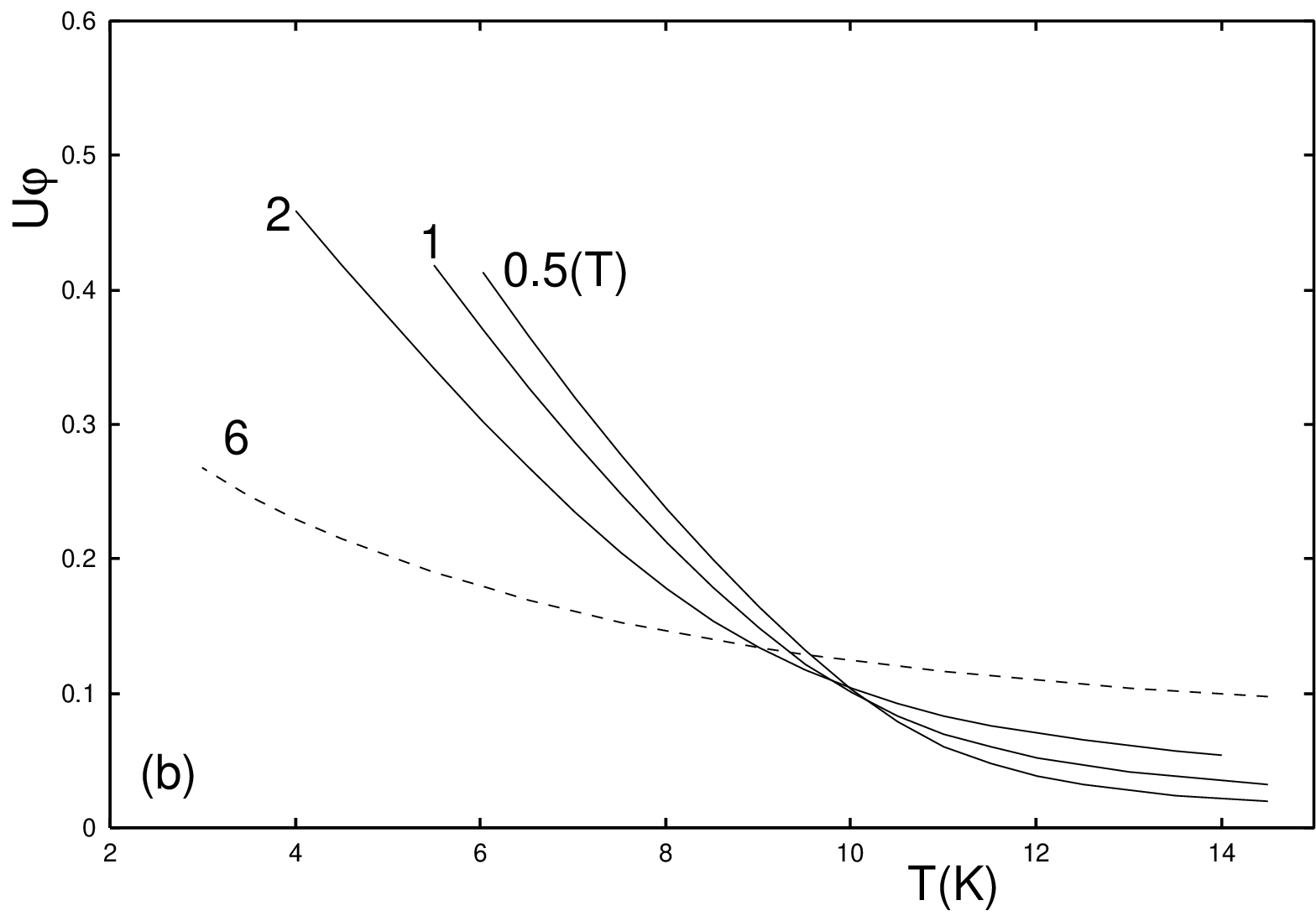}}
\caption{(a) Fitting results to resistivity data \cite{Sasaki} of $\kappa$-(ET)$_2$Cu[N(CN)$_2$]Br and (b) the expected $U_\phi$-curves corresponding to the solid curves in (a). For instance, $T^*_{c2}(H=6({\rm T}))$ lies near $8$(K).}
\label{} 
\end{figure}

Resistivity data of $\kappa$-(ET)$_2$Cu[N(CN)$_2$]Br \cite{Sasaki} are fitted in terms of the present theory, and the results are shown in Fig.9 (a). The 3D form of $\sigma_{\rm vg}$, eq.(17), is assumed together with $t_{\rm vg}=(1-H/H_0)^2/(1+3.52(H/H_0)^{1/2})$ and $c_p=0.0035$, and the material parameters used there are $\lambda(0)=0.72$($\mu$m), $T_0 \simeq T_{c0}=12$ (K), $H_0=18.3$(T), $d=1.5$(nm), and the normal resistance $R_n$($\Omega$)$=0.0574+0.0243 \, t^{5/2}$. As a result of the large $\lambda(0)$-value, the quantum SC fluctuation becomes essential with increasing $H$ and leads to high field curves following $R_n(T)$ even in $T \ll T_{c0}$. Further, the $c_p$-value and the $t_{\rm vg}$-form suggest a much weaker pinning effect than in LSCO. For comparison, the $U_\phi$ curves obtained in terms of the same set of parameters are shown in Fig.9 (b). Note that, within the LLL, $U_\phi$ is equivalent to the magnetization, and hence that Fig.9 (b) can be also regarded as a typical example of thermodynamic quantity. A crossing behavior just below $T_{c0}$, which is familiar through magnetization data in many optimally-doped cuprate materials, is seen in lower fields below 2(T). It implies that the fluctuation property below 2(T) is purely thermal. 
In contrast, the $U_\phi(T)$-curve in 6(T) deviates from the crossing behavior in lower fields and is anomalously broadened, reflecting the stronger (quantum) fluctuation. Such an additional broadening of thermodynamic quantities due to the quantum fluctuation is an opposite trend to the corresponding behavior near $T=0$, where a broadening of such quantities diminishes reflecting a rise of dimensionality of SC fluctuation in the quantum regime near $T=0$ \cite{RI4}. This is why, as mentioned in $\S 1$, the quantum behavior at high temperatures should be distinguished from that near $T=0$. The above features of $U_\phi$ seem to be consistent with the magnetization data in ref.18.
\begin{figure}[t]
\scalebox{0.3}[0.3]{\includegraphics{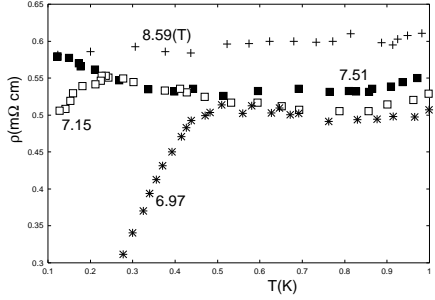}}
\caption{Data of resistivity of $\kappa$-(ET)$_2$Cu(NCS)$_2$ near $T=0$ and in strong fields \cite{Sasaki2}.}
\label{} \end{figure}

\begin{figure}[t]
\scalebox{0.3}[0.3]{\includegraphics{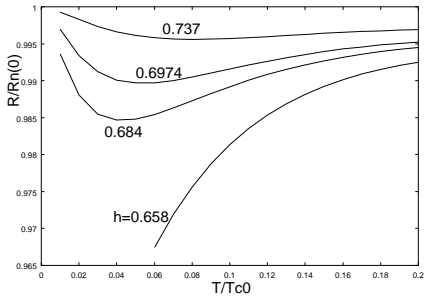}}
\caption{Resistivity curves computed for a qualitative comparison with the data in Fig.10.}
\label{} \end{figure}

Next, the resistive behavior near $T=0$ will be discussed based on the data of  $\kappa$-(ET)$_2$Cu(NCS)$_2$ shown in Fig.10 which corresponds to Fig.3 of ref.49. In the highest field 8.59(T) above $H_0$ defined through a study of Shubnikov-de Haas effect \cite{Sasaki2}, $\rho(T)$ shows a $T$-dependence indicative of a metallic normal state and approaches a residual value $\rho_n(0)$. In lower fields just below $H_0$, $\rho(T)$ first decreases upon cooling as a result of thermal SC fluctuation, while it begins to increase on further cooling and approaches $\rho_n(0)$ (The ultimate drop of $\rho$ due to the VG fluctuation at a low $T$ of resistance below 7(T) will not be considered here). This insulating behavior occurs even much below $H_0$ and hence, is a phenomenon of a SC origin. This feature in Fig.10 is a direct evidence of an insulating behavior \cite{RI4,RI10} arising from a purely dissipative quantum SC fluctuation in cases with metallic normal resistance. 
For comparison, we give in Fig.11 examples of computed $\rho(T)$ curves with such a fluctuation-induced insulating behavior at low enough $T$. In Fig.11, the 2D $\sigma_{\rm vg}$, eq.(18), was used, and the parameter values, $\lambda(0)=1.1$($\mu$m), $b_p=(2 \pi \xi_0^2 d) \, 0.06$, $T_0=T_{c0}=25$(K), and $H_0=19$(T), were chosen. Further, the relations $(R_Q d \sigma_n)^{-1}=0.35(1+0.05(T/T_{c0})^{5/2})$ and $\mu_0=T/T_{c0} - 1 +H/H_0$ were used, and, for simplicity, the vertex correction to the pinning strength was neglected by setting ${\tilde b}_p=b_p$. 

\section{$s$-wave Dirty Films}
In ref.23, we have proposed a theory of field-tuned superconductor-insulator (FSIT) behavior in {\it homogeneously} disordered thin SC films with $s$-wave pairing on the basis of a familiar electronic model in dirty limit including effects of a repulsive mutual interaction between electrons. It has been argued there how, reflecting differences in $T$-dependences between various components in $\sigma$, the resistance value $R_c$ on an apparent FSIT field $B_c^*$ and the resistive behavior around $B_c^*$  are affected by the normalized value $R_n/R_Q$ of the (high temperature) sheet resistance. However, no detailed computation results based on the derived GL action were given there. Motivated by a recent finding on  $R_c$ v.s. $R_n$ relation \cite{Rochester}, some computed results on resistance curves consistent with the experimental observations \cite{Rochester,Phillips} will be presented here. In contrast to the previous two sections, we take account of the fact that the resistance data of $s$-wave amorphous films are conventionally discussed in terms only of the $R_n/R_Q$ value and will not try here to fit to real data. Actually, the expressions (see Appendix A) we use here for microscopic parameters were derived from the simplest extension of the BCS model to the case with both the disorder and a repulsive interaction between the electrons and may not explain quantitatively materials, for example, with a strong spin-orbit scattering. 
\begin{figure}[t]
\scalebox{0.4}[0.4]{\includegraphics{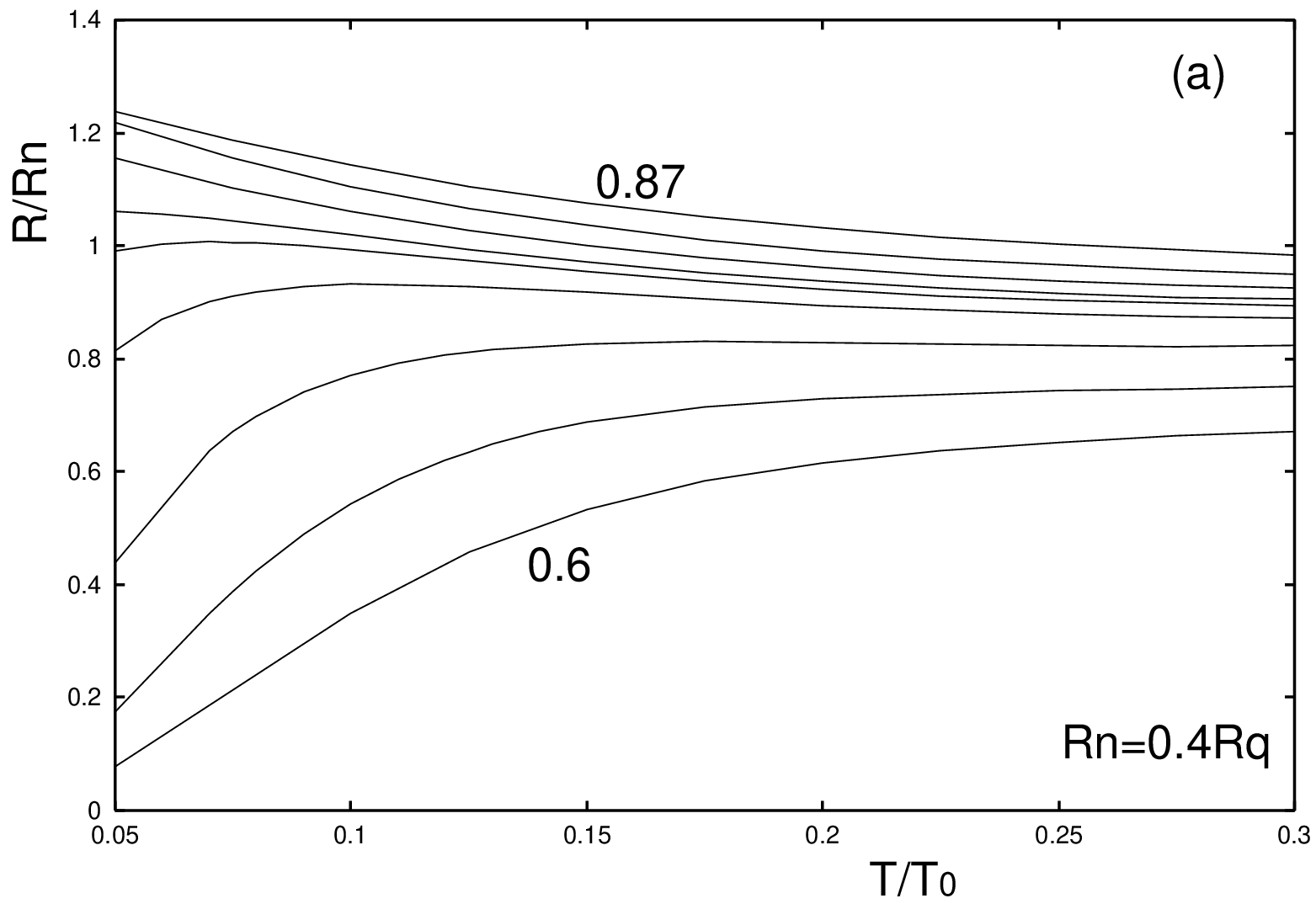}}
\scalebox{0.4}[0.4]{\includegraphics{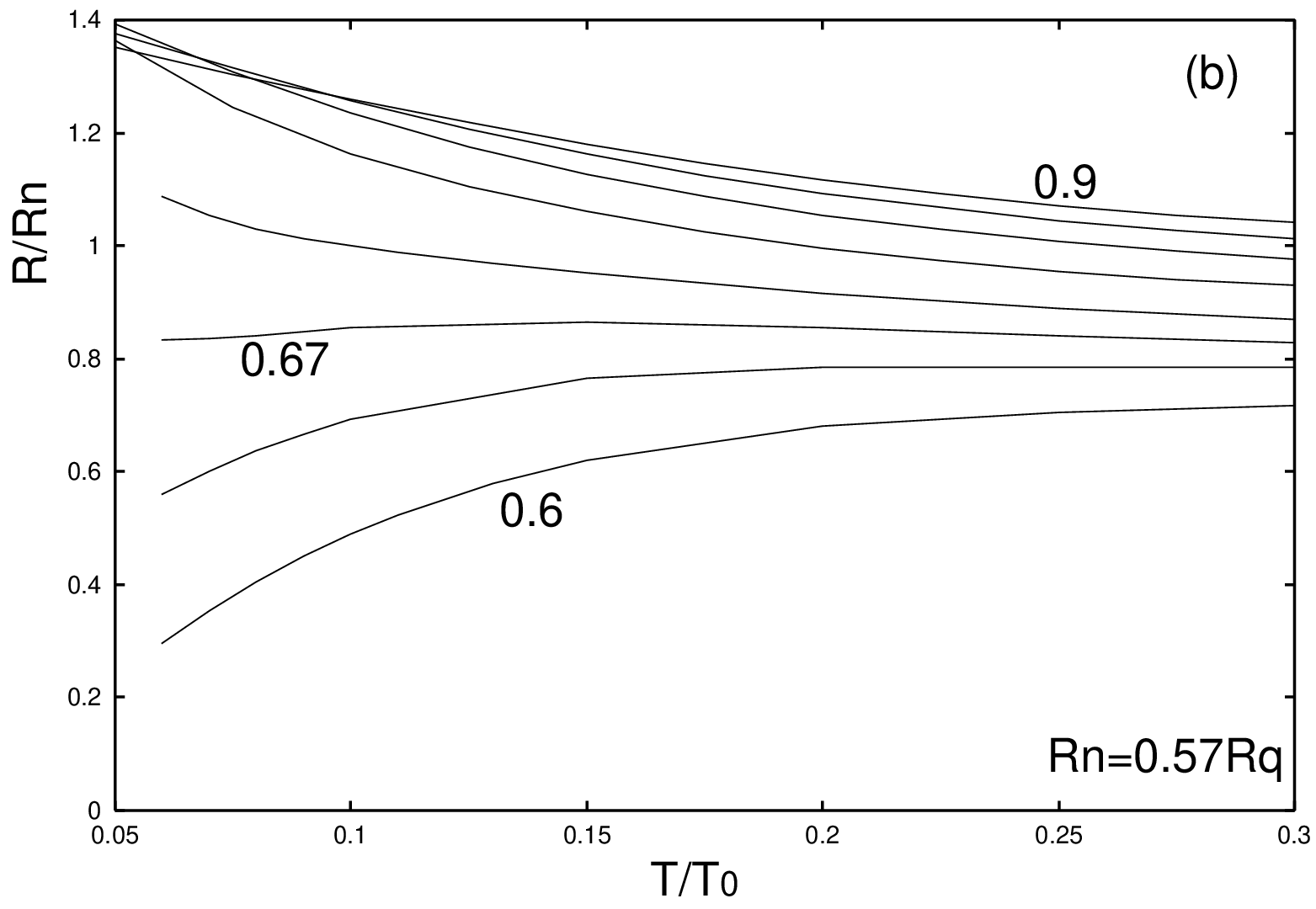}}
\scalebox{0.4}[0.4]{\includegraphics{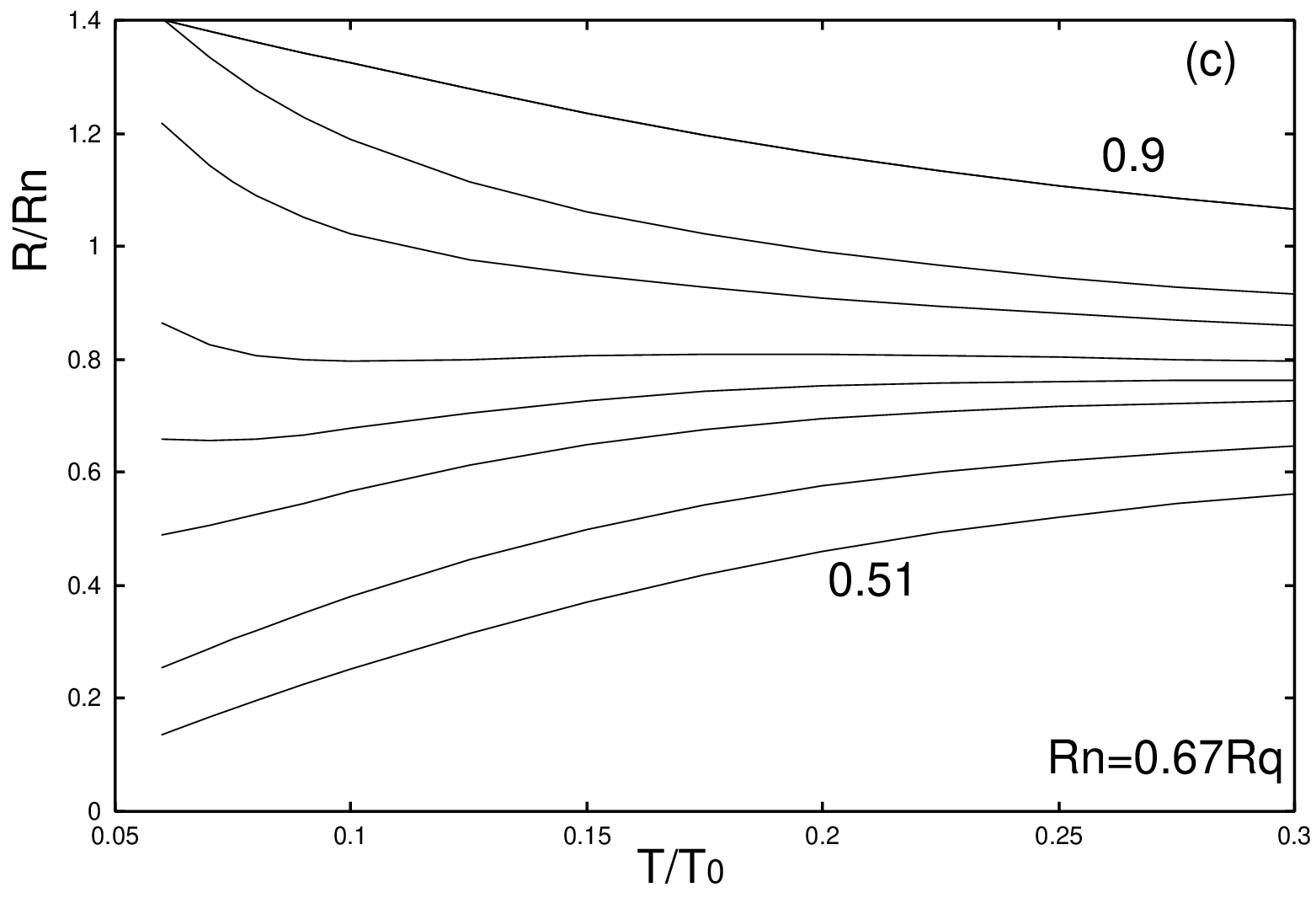}}
\caption{Calculated resistivity curves imagining $s$-wave dirty films for three $R_n/R_Q$ values : (a) $R_n/R_Q=0.4$ in $H/H_0= 0.87$, $0.82$, $0.79$, $0.77$, $0.76$, $0.74$, $0.7$, $0.65$, and $0.6$, (b) $R_n/R_Q=0.57$ in $H/H_0 = 0.9$, $0.85$, $0.8$, $0.75$, $0.7$, $0.67$, $0.64$, and $0.6$, and (c) $R_n/R_Q=0.67$ in $H/H_0=0.9$, $0.8$, $0.67$, $0.63$, $0.61$, $0.59$, $0.55$, 
and $0.51$. }
\label{} \end{figure}
\begin{table}
\begin{tabular}{ccccccc}
$R_n/R_Q$ &0.25 &0.4 &0.45 &0.57 &0.625 &0.67 \\
\hline
$B_c^*/H_0$ &0.8 &0.76 &0.72 &0.68 &0.64 &0.61 \\
$R_c/R_n$ &0.98 &0.98 &0.92 &0.83 &0.75 &0.65 \\
\end{tabular}
\caption{\label{tab:table3}$R_n/R_Q$ dependences of the normalized critical resistance $R_c/R_n$ and of the normalized critical field $B_c^*/H_0$, estimated for each case of the calculated resistivity curves illustrated in Fig.12.}
\end{table}

The coefficients of each term of the GL action for the $s$-wave SC films were studied elsewhere \cite{RI11,RI12}. In Appendix A, they will be given in a form useful for numerical computations. 
Applying those GL coefficients to eq.(9) and using eq.(18) as $\sigma_{\rm vg}$, we have examined the resistance curves near $B_c^*$ in details, and the results are shown in Fig.12. Since, as mentioned in Appendix A, the value of the parameter $\tau T_0$ was fixed, the normalized sheet resistance (defined at high $T$) $R_n/R_Q$ is the only material parameter for our calculation of $R_Q d \sigma$ and determines {\it both} strengths of quantum fluctuation and vortex pinning. Let us define $B_c^*$ as the field at which $R(T)$ takes a value $R_c$ insensitive to $T$ at low enough $T$ within the examined 
temperature range. 

From the resistance curves shown in Fig.12, one will find the 
following two features on the curves near $B_c^*$. 
First, for lower values of $R_n/R_Q$, the curves near $B_c^*$ show an 
insulating $T$-dependence at intermediate temperatures, while they, for large $R_n/R_Q$ values, rather decrease upon cooling in the same temperature region. 
This feature has been seen in various data \cite{Hebard,Valles,Gant,Goldman} and justifies the scenario predicted in ref.23. At least within the present model of $s$-wave thin films, the decrease of $R(B_c^*)$ upon cooling for large enough $R_n/R_Q$-values is dominated by an enhancement of pinning strength, while its increase in smaller $R_n$-values occurs primarily reflecting an enhancement of quantum SC fluctuation. For instance, the insulating $R$($B \simeq B_c^*$) for $R_n = 0.4 R_Q$ is controlled primarily by the quantum behavior of $\sigma_{\rm fl}$ rather than $\sigma_{\rm vg}$. Next, as listed in Table III, $R_c$ at lower $R_n$ ($ \leq 0.5R_Q$) values roughly coincides with $R_n$, while $R_c$ in $R_n \geq 0.6$ rather decreases with increasing $R_n$. The relation $R_c \simeq R_n$ at low disorder (lower $R_n$) coincides with the experimental results in $R_n \leq 5$(k$\Omega$) summarized in Fig.4 of ref.52, while the saturation and reduction of $R_c$ at higher disorder qualitatively agree with Fig.4 of ref.51. 

Finally, we will comment on the validity of eq.(9) used in obtaining Fig.12 on the basis of the microscopic informations in Appendix A. When replacing eq.(4) with eq.(9), the pinning effect was underestimated. This simplification needs to be reconsidered in a close vicinity of a VG transition point (i.e., $\mu_{\rm vg} \to +0$). In trying to fit various data in $\S 3$ and $4$, we did not have to discuss the details of resistivity data in the vicinity of the VG 
transition. In contrast, describing in details the resistive behaviors in 2D disordered thin films at low enough $T$ and near $B_c^*$ (i.e., near the quantum VG transition) is needed to clarify the physics of FSIT. We have partly carried out numerical computations based on eq.(4), although such a cumbersome analysis will not be presented here, and found that, roughly speaking, numerical results following from eq.(9) in the cases of $R_n/R_Q$-values used in Fig.12 are quantitatively unreliable only in $T/T_{c0} < 0.05$. For this reason, no data in $T/T_{c0} < 0.05$ have been shown in Fig.12. 

\section{Comments and Discussion} 
First, a view extended over a wider temperature range of the resistivity curves in Fig.12 (b) is shown in Fig.13. From this figure, a $T=0$ 2D VG transition point $B_c^*$ is suggested to apparently lie much below $H_{c2}^*(0)$, because $B_c^*$ itself is lowered by the quantum SC fluctuation \cite{RI11}, while the FSIT behavior appearing around $B_c^*$ is a reflection of the VG fluctuation. Since the flat or insulating resistive behaviors appear in $H \geq B_c^*$, the temperature at which $R(T)$ in a field slightly below $B_c^*$  
drops {\it inevitably} lies far below $T_{c2}^*(H)$ indicated by a dark circle in the figure. Since the FSIT behavior is a consequence of strong quantum SC fluctuation, this example means that the sharp resistive drop much below $H_{c2}^*(T)$ in systems with strong quantum fluctuation is not an artifact of approximations used in calculations but a generic feature occurring commonly in clean and dirty limits irrespective of calculation methods of $R(T)$. 
\begin{figure}[t]
\scalebox{0.4}[0.4]{\includegraphics{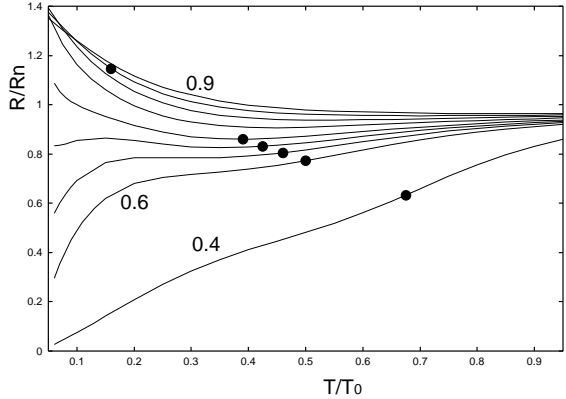}}
\caption{Extended view of Fig.12 (b) to higher temperatures. Here, a curve in $H=0.4 H_0$ is added further. The darked circle on each $\rho(T)$ curve denotes $T_{c2}^*(H)$ in each $H$.}
\label{} \end{figure}

As shown in $\S 3$, the quantum SC fluctuation plays important roles in many cuprate materials under a high field. If the quantum SC fluctuation becomes essential, as in the example of Fig.1, due to a growth of $\lambda(0)$, one might expect the critical region of the thermal SC transition in $H=0$ to also widen due to the $\lambda(0)$-growth. Actually, it was questioned in ref.26, through Figs.7 and 8 there, whether the pictures arguing a pseudogap of SC fluctuation origin are consistent with the fact that the critical region of the $H=0$ transition does not widen much with underdoping. If the approach in $\S 2$ is extended to the $H=0$ case, however, we find two theoretical facts satisfactorily explaining the sharp $H=0$ transition in underdoped cuprates. First, it is easily noticed that the microscopic $T$-dependence in the broad SC pseudogap regime of the coefficient of the gradient term in GL model results in a reduction of the width of the critical region around $T_{c0}$ in $H=0$. Actually, by noting that this GL coefficient in $H=0$ is nothing but that of the $H$-linear term in eq.(23), one easily finds that the 2D Ginzburg number {\it near} $T_{c0}$ is given, {\it consistently} with eq.(2), by $16 \pi^2 [\lambda(0)]^2 k_B (T_{c0})^3/(\phi_0^2 d T_0^2)$, where the reduction factor $(T_{c0}/T_0)^2$ arises from a {\it microscopic} $T$-dependence above $T_{c0}$ of the gradient term. Within the present approach in $\S 2$ where no specific origin leading to a reduction of the time scale $\gamma$ is assumed, this may be adequate for understanding the unexpectedly narrow \cite{Wang1,Dima} $H=0$ critical region of underdoped cuprates. In a system with strong fluctuation, we have another mechanism of a shrinkage of the $H=0$ critical region due to the quantum SC fluctuation itself. An explanation of this mechanism will be given in Appendix B. This mechanism is particularly effective when the enhancement of quantum SC fluctuation is primarily due to a decrease of the dissipative time scale $\gamma$. In any case, a shrinkage of the $H=0$ critical region in a system with strong fluctuation is not a surprising phenomenon. 

We need to comment here on the definition of penetration depth comparable with experimental 
data \cite{Pana}. The actual penetration depth in $H=0$ is defined as the mass of the gauge field through the gradient term. Then, by taking account of the $T$-dependence of the gradient term in the SC pseudogap regime mentioned above, the penetration depth to be observed near $T_{c0}$ is found to be not $\lambda(0)/(1-T/T_{c0})^{1/2}$ but $T_{c0} \lambda(0)/[T_0 (1 - T/T_{c0})^{1/2}]$. 
If a $T$-dependence of the coefficient $b$ in the SC pseudogap region can be 
neglected, the former is the quantity to be observed as the penetration depth 
in $H=0$ and at low $T$ where the fluctuation is negligible. Thus, in underdoped materials with wider SC pseudogap region, the $[\lambda(T)]^{-2}$ v.s. $T$ curve is not linear even approximately, and, as observed in ref.59, the local slope $|d \lambda^{-2}/dT|$ increases on approacing $T_{c0}$ from below. However, we note that the $\lambda(0)$-values we have estimated by fitting are slightly longer than those estimated experimentally \cite{Pana}. If effects of the competing orders may be relatively negligible, the $b$-value near $T_{c0}$ will be, as well as $\gamma_0$, larger than that near $T_0$. Thus, the differences between $\lambda(0)$-values obtained through Fig.6 and their experimental data \cite{Pana} do not necessarily require modification of the present theory.  

It should be stressed that key data \cite{Ito,Capan1,Wang2,Capan2} showing the {\it uncorrelated} \cite{RI7} behavior between the Nernst coefficient (or the magnetization) and the resistivity were explained in the preceding sections without taking account of electronic states peculiar to materials near a Mott transition. In the case of cuprates, such a behavior tends to arise in systems with low $H_c(0)$ such as the overdoped and underdoped materials, while it is rarely seen in optimally-doped cuprates with higher $H_c(0)$ \cite{Mac2,Mac3}. This {\it nonmonotonic} doping dependence in the cuprates we have explained in $\S 3$ strongly suggests that the {\it uncorrelated} behavior is not a consequence of a strong reduction of the friction coefficient due to a microscopic mechanism \cite{Dima,Lee2} peculiar to systems close to a Mott transition. It does not seem to us that the observation \cite{HS,Mac2,Mac3}, that the uncorrelated behavior and the sharp resistive drop are more remarkable in higher fields, while the fan-shaped broadening is usually seen in lower fields, can be explained in terms of such an approach \cite{Dima,Lee2} based on a single vortex picture valid in low enough fields. Further, as emphasized in $\S 1$ and elsewhere \cite{RI7}, the uncorrelated resistive behavior is {\it not} peculiar to the cuprates near a Mott transition but also appears commonly in other superconductors such as the organic materials and disordered films with an $s$-wave pairing (see Figs.9 and 13). 

In Fig.6 (a), we have included a $\rho(T)$ curve at 29(T) which is above $H_{c2}^*(0)$ (see also Fig.8). The sharp drop of this curve near 10(K) indicates a 3D VG transition point {\it above} $H_{c2}^*(T)$. Since, in principle, the VG transition may occur as far as the SC fluctuation is present, such a 3D VG transition and hence, a sharp resistive drop {\it in} the SC pseudogap region $T_{c2}^*(H) < T < T_0(H)$ may be present, in contrast to the scenario in ref.60 for the data \cite{Mac}, even in {\it homogeneously} disordered materials. The features seen in (K,Ba)BiO$_3$ reported in ref.62 may be a remarkable example with $T_{c2}^* < T_{\rm vg} < T_0(H)$. 

It is interesting to connect the nonmonotonic \cite{Wang1} doping dependence of $T_\nu(x)$ in LSCO to a sign change \cite{Matsu} of the fluctuation Hall conductivity $\sigma_{{\rm fl}, xy}$. It is now clear \cite{Kokubo,Ott} that a sign reversal of Hall conductivity, usually seen in the vortex liquid regime, may occur above $H_{c2}^*(T)$, depending on the materials. It means that this Hall-sign reversal should be understood based on the fluctuation scenario \cite{RI13} unrelated to the electronic states of vortex cores. According to this scenario, the sign of $\sigma_{{\rm fl},xy}$ is determined by that of $\partial T_c^{\rm MF}/\partial x$, and hence, if $T_\nu$ is essentially identical with $T_c^{\rm MF}$, the sign change of $\partial T_\nu/\partial x$ should appear directly as that of $\sigma_{{\rm fl},xy}$, where $T_c^{\rm MF}$ was defined in $\S 3$. According to ref.63, the LSCO data in $x \geq 0.12$ show a Hall-sign reversal, while available $\rho_{xy}$ data in $x=0.08$ \cite{Lang} have not shown any sign reversal, consistently with our expectation. On the other hand, according to ref.15, it appears that the Hall data of very underdoped Bi2201 still show a sign reversal, consistently with the monotonic doping dependence of $T_\nu$ of this material \cite{Wang1}. This is why the identification between $T_\nu$ and $T_c^{\rm MF}$ seems to be consistent with available Hall-resistance 
data.

In hole-doped cuprates, the positive magnetoresistivity in the SC pseudogap region above $T_{c0}$ is enhanced with underdoping. Actually, this trend has led the authors in ref.47 to argue that $\xi_0$ increases with underdoping. This enhanced magnetoresistivity above $T_{c0}$ can be seen as having a common origin to the resistivity curves \cite{Capan1,Wang2,Capan2} following the extrapolated normal curve even much below $T_{c2}^*(H)$ ($< T_{c0}$). Actually, in a strong SC fluctuation case such that its quantum nature is no longer negligible, the Gaussian approximation for the SC fluctuation fails, and the interaction between the SC fluctuations may be important even much above $T_{c0}$. Hence, the quantum SC fluctuation may be a 
main origin of the enhanced \cite{Ando} magnetoresistance in the SC pseudogap 
region (i.e., $T_{c0} < T < T_0$). In any case, as demonstrated in ref.20, focusing only on $\rho(T)$ data in cases with low $H_c(0)$ and hence, with large quantum fluctuation, tends to lead to an erroneous estimation of material parameters. A simultaneous study of other quantities, such as $U_\phi$, measured consistently with $\rho$ is indispensable. Further, the neglect of SC pseudogap region in thermodynamic quantities such as the magnetization \cite{Li} has also erroneously led one to concluding a $\xi_0$ increasing with underdoping. 

During preparing this manuscript for submission, several related works \cite{Wang3,md} on transport phenomena in the vortex liquid regime were reported. The {\it monotonic} decrease \cite{RI7} of $\xi_0$ accompanying underdoping we have concluded through fitting to data was also argued \cite{Wang3} from an extrapolation of Nernst data to very higher fields in which experimental measurements cannot be performed. The uncorrelated behavior between the resistivity and the Nernst data seen in underdoped Pr$_{2-x}$Ce$_x$CuO$_4$ was erroneously interpreted in ref.74 as an insensitivity of the resistivity to SC fluctuations. As we have clarified here, a theoretically valid explanation is provided only by noting the reduction of $\sigma_s$ induced by the quantum SC fluctuation at nonzero temperatures.

\begin{acknowledgements}
The author thanks T. Sasaki, C. Capan and W. Lang for providing him their unpublished data. This work was partly supported by a Grant-in-Aid from the Ministry of Education, Sports, and Culture, Japan.
\end{acknowledgements}

\appendix

\section{}

Here, we give numerically useful expressions of the coefficients of the GL action in the case of $s$-wave dirty films \cite{RI5,RI11,RI12}. In obtaining them, we are largely based on the ordinary dirty limit in quasi 2D case where the 2D diffusion and fluctuation propagators are assumed consistently with a 3D electronic state (see the second paper of ref.23), and effects of Coulomb interaction between the electrons are included perturbatively and modelled in a form interpolating between the low $T/H$ and high $T/H$ regions. The GL coefficients of the quadratic term we use in $\S 5$ are given by
\begin{eqnarray}
\mu_0={\rm ln}(T/T_0) + \sum_{n \geq 0} \frac{1}{(n+1/2)((n+1/2){\tilde t}+1)} \;
\\ \nonumber
 + \frac{R_n}{24 \pi R_Q} (\, {\rm ln}(2 T_0
 /(T + T_{cr}^{\rm mf})) \, )^3, 
\end{eqnarray}
\begin{eqnarray}
({\cal G}_1(0))^{-1} \simeq \mu_1- \mu_0 \;
\\ \nonumber
= 2 {\tilde t} \sum_{n \geq 0} \frac{1}{(1+{\tilde t}(n+1/2))(3+{\tilde t}(n+1/2))}, 
\end{eqnarray}
and 
\begin{eqnarray}
\gamma_0 \simeq \gamma_1=\gamma^{(0)}/(1 + \frac{5 R_n}{8 \pi R_Q} ({\rm ln}(2 T_0/(T+T_{cr}^{\rm mf})) \, )^2 ) \; 
\\ \nonumber 
= \frac{{\tilde t}}{4 \pi T_{cr}^{\rm mf}} \sum_{n \geq 0} (1+{\tilde t}(n+1/2))^{-2} \;
\\ \nonumber
\times (1 + \frac{5 R_n}{8 \pi R_Q} ({\rm ln}(2 T_0/(T+T_{cr}^{\rm mf})) \, )^2 )^{-1}, 
\end{eqnarray}
where $T_{cr}^{\rm mf}=0.14 T_0 H/H_{c2}^d(0)$, $H_{c2}^d(0)$ is $H_0$ in dirty limit, ${\tilde t}=T/T_{cr}^{\rm mf}$, $R_n$ is the sheet resistance of a quasi 2D film at high $T$ inversely proportional to the film thickness $d$, and $\gamma_0$ in dirty limit with no electron repulsion was denoted as $\gamma^{(0)}$. For simplicity, a SC pseudogap region is assumed to be absent (i.e., $T_0=T_{c0}$), and $\gamma_0 = \gamma_1$ was additionally assumed because a difference between them is less important in dirty limit than that in clean limit (see eq.(23)) where $\gamma_1$ tends to vanish in $T \to 0$ limit. No interaction correction to ${\cal G}_1(0)$ was included because, as explained in $\S 2$ and ref.23, a detailed form of ${\cal G}_1(0)$ is {\it not} reflected in $\sigma_{\rm fl}$ and $\sigma_{\rm vg}$. 

On the other hand, the coefficients $b$ and $b_p$, respectively, of the interaction and pinning terms will be expressed as 
\begin{equation}
b=\frac{r_B^2 d R_n}{3 R_Q} \frac{{\tilde t}}{T_{cr}^{\rm mf}} \sum_{n \geq 0} (1+{\tilde t}(n+1/2))^{-3},
\end{equation}
and 
\begin{eqnarray}
b_p=\biggl(\frac{r_B R_n}{R_Q} \biggr)^2 \frac{d {\tilde t}}{6 \pi} \sum_{n \geq 0} \sum_{m \geq 0} (n+m+1)^{-1} \;
\\ \nonumber
\times (1+{\tilde t}(n+1/2))^{-1}  (1+{\tilde t}(m+1/2))^{-1}. 
\end{eqnarray}
Effects of electron-repulsion on $b$ and $\gamma_0$ were included altogether just in $\gamma_0$ since its effect on $\gamma_0$ was estimated in ref.53 to be much larger than that in $b$, and they usually appears as the quantum fluctuation strength $\propto b/\gamma_0$ in the quantum regime. 
We note that $b_p$ is $U_p f_{00}(0)$ in the notation of ref.23. 
Through the computations of resistivity curves shown in Figs.12 and 13, for simplicity, another parameter $2 \pi T_0 \tau$, where $\tau$ denotes the elastic scattering time, was fixed to 0.5. We expect a detailed value of this parameter not to significantly affect the numerical results. 

As a normal conductivity $\sigma_n$ for this case, the expression 
\begin{equation}
R_Q d \sigma_n = (1 - \frac{R_n}{3 R_Q} 
\frac{2 \pi r_B^2 d \gamma^{(0)}}{b} \Sigma_0)/[1+\frac{R_n}{4 \pi R_Q} {\rm ln}(T_0/T)]
\end{equation}
was used in the following figures, 
where $\Sigma_0$ is given by eq.(10). The second term of the numerator corresponds to the low $T$ form of additional quantum SC fluctuation contributions \cite{GaL1} to $\sigma_n$ excluded in the GL approach and was argued \cite{RI5,GaL2} to be the origin of the {\it fluctuation-induced} \cite{Okuma} negative magnetoresistance in 2D and at low $T$. Such a behavior visible in higher $H$ and lower $T$ in Fig.12 is a consequence of this 
fluctuation term. 
Further, to represent ${\rm ln}(T_0/T)$-term in the denominator of $\sigma_n$ ensuring $\sigma_n(T \to 0) \to 0$, a form of $\sigma_n$ expected \cite{LR} in the case with a strong spin-orbit scattering and with long-ranged Coulomb interaction was conveniently assumed as a model. In this way, when $H$ and $T$ are scaled, respectively, by $H_{c2}^d(0)$ and $T_0$, both the GL coefficients and $R_Q d \sigma_n$ are parametrized only by $R_n/R_Q$.

\vspace{5mm}

\section{}

Here, it will be explained how the quantum fluctuation itself reduces the width of critical region of the $H=0$ transition at $T_c$. To sketch the essence of this effect, we will just treat a counterpart of eq.(4) (with $b_p=\Delta \Sigma_l=0$) in the Hartree approximation and in $H=0$. 
In the isotropic 3D case, the renormalized mass $\mu_R$ (corresponding to $[ {\cal G}_0(0) ]^{-1}$) yields the relation 
\begin{equation}
\mu_R- {\rm ln}(T/T_c) = \sum_\omega [ \zeta(\mu_R; \omega) 
- \zeta(0; \omega) ],
\end{equation}
where 
\begin{equation}
\zeta(\mu_R; \omega)= 2 \pi [\varepsilon_G^{(3)}(T)]^{1/2}
\int_{\bf q} \frac{1}{\mu_R + q^2 + \gamma|\omega|},
\end{equation}
$\varepsilon_G^{(3)}(T) = [{16 \pi^2 (\lambda(0))^2}/(\beta \xi_0 \phi_0^2)]^2$ is the 3D Ginzburg number at temperature $T$, and $T_c$ was redefined so that $\mu_R(T=T_c)=0$ is satisfied. The thermal ($\omega=0$) part of the r.h.s. of eq.(B1) becomes $ - (\varepsilon_G^{(3)}(T))^{1/2} \sqrt{\mu_R}$. If keeping only this contribution in the r.h.s. and focusing on the low $\mu_R$ limit, the critical exponent $\nu=1$ of the correlation length in the spherical limit is obtained. On the other hand, the sum of other $\omega \neq 0$ terms in the r.h.s. is well approximated by $- \pi^{-2} \mu_R \sqrt{\varepsilon_G^{(3)}(\gamma^{-1})}$ if the quantum fluctuation is strong enough. Then, solving eq.(B1) with respect to $\mu_R$, one finds the width of critical region to be estimated as $T_c \varepsilon_G^{(3)}(T_c)/[16 (1+\pi^{-2} \sqrt{\varepsilon_G^{(3)}(\gamma^{-1})})]$, implying that, with decreasing $\gamma$, the critical region is narrower. The basic reason of this narrowing of the critical region due to the quantum fluctuation consists in the fact that the width of quantum critical region at $T=0$ and in 3D is negligible, because the dimensionality of the dissipative quantum critical fluctuation at $T=0$ in $D$-dimension is $D+2$, which is above the upper critical dimension (i.e., four) for $D > 2$ \cite{Millis}.

\vfil\eject


\begin{thebibliography}{99}
\bibitem{RI1} R. Ikeda, T. Ohmi and T. Tsuneto: J. Phys. Soc. Jpn. {\bf 60} (1991) 1051. 
\bibitem{RI2} R. Ikeda, T. Ohmi and T. Tsuneto: Phys. Rev. Lett. {\bf 67} (1991) 3874. 
\bibitem{Sarti} S. Sarti, D. Neri, E. Silva, R. Fastampa and M. Giura: Phys. Rev. {\bf B} 56 (1997) 2356. 
\bibitem{RI3} R. Ikeda: J. Phys. Soc. Jpn. {\bf 70} (2001) 219. 
\bibitem{FFH} D. S. Fisher, M. P. A. Fisher and D. A. Huse: Phys. Rev. B {\bf 43} (1991) 130. 
\bibitem{Natter} T. Nattermann and S. Sheidl: Adv. Phys. {\bf 49} (2000) 607. 
\bibitem{Mac} A. P. Mackenzie, S.R. Julian, G.G. Lonzarich, A. Carrington, S.D. Hughes, R.S. Liu and D.C. Sinclair: Phys. Rev. Lett. {\bf 71} (1993) 1238. 
\bibitem{HS} M. Suzuki and M. Hikita: Phys. Rev. B {\bf 44} (1991) 249. 
\bibitem{Carrington} A. Carrington, A. P. Mackenzie and A. Tyler: Phys. Rev. 
B {\bf 54} (1997) R3788. 
\bibitem{Naito} S. Kleefisch, B. Welter, A. Marx, L. Alff, R. Gross and M. Naito: Phys. Rev. B {\bf 63} (2001) 100507; F. Gollnik and M. Naito: Phys. Rev. B {\bf 58} (1998) 11734. 
\bibitem{Mac2} D. J. C. Walker, O. Laborde, A.P. Mackenzie, S.R. Julian, A. Carrington, J.W. Loram and J.R. Cooper: Phys. Rev. B {\bf 51} (1995) 9375. 
\bibitem{Mac3} A. Carrington, A.P. Mackenzie, D.C. Sinclair and J.R. Cooper: Phys. Rev. B {\bf 49} (1994) 13243. 
\bibitem{Gan} V. F. Gantmakher, G.E. Tsydynzhapov, L.P. Kozeeva and A.N. Lavrov: JETP {\bf 88} (1999) 148. 
\bibitem{Karpinska} K. Karpinska, A. Malinowski, M.Z. Cieplak, S. Guha, S. Gershman, G. Kotliar, T. Skoskiewicz, W. Plesiewicz, M.Berkowski and P. Lindenfeld: Phys. Rev. Lett. {\bf 77} (1996) 3033. 
\bibitem{Vedeneev} S. I. Vedeneev, A.G.M. Jansen, E. Haanappel and P. Wyder: Phys. Rev. B {\bf 60} (1999) 12467. 
\bibitem{Capan1} C. Capan, K.Behnia, J.Hinderer, A.G.M. Jansen, W. Lang, C. Marcenat, C. Marin and J. Flouquet: Phys. Rev. Lett. {\bf 88} (2002) 056601. 
\bibitem{Wang2} Y. Wang, N.P. Ong, Z.A. Xu, T. Kakeshita, S. Uchida, D.A. Bonn, R. Liang and W.N. Hardy: Phys. Rev. Lett. {\bf 88} (2002) 257003. 
\bibitem{Sasaki} T. Sasaki et al.: unpublished. See also A. Matsuyama, T. Sasaki, T. Fukase and N. Toyota: Physica C {\bf 263} (1996) 534. 
\bibitem{Ito} H. Ito, T. Ishiguro, T. Komatsu, G. Saito and H. Anzai: Physica B {\bf 201} (1994) 470. 
\bibitem{RI7} R. Ikeda: Phys. Rev. B {\bf 66} (2002) 100511(R). 
\bibitem{RI4} R. Ikeda: Int. J. Mod. Phys. B {\bf 10} (1996) 601. 
\bibitem{Huebner} H. C. Ri, R. Gross, F. Gollnik, A. Beck, R.P. Huebener, P. Wagner and H. Adrian: Phys. Rev. B {\bf 50} (1994) 3312. 
\bibitem{RI5} H. Ishida and R. Ikeda: J. Phys. Soc. Jpn. {\bf 71} (2002) 254 
and (2002) 2352. 
\bibitem{Blatter} G. Blatter and B. Ivlev: Phys. Rev. Lett. {\bf 70} (1993) 
2621. 
\bibitem{RI6} R. Ikeda: Physica B {\bf 329-333} (2003) 1457.
\bibitem{Wang1} Y. Wang, Z.A. Xu, T. Kakeshita, S. Uchida, S. Ono, Y. Ando and N.P. Ong: Phys. Rev. B {\bf 64} (2001) 224519. 
\bibitem{Smith} D. R. Niven and R. A. Smith: Phys. Rev. B {\bf 66} (2002) 
214505. 
\bibitem{Tesanovic} O. Vafek, A. Melikyan and Z. Tesanovic: Phys. Rev. B {\bf 64} (2001) 224508. 
\bibitem{Matsuda} K. Izawa, A.Shibata, H.Takahashi, Y.Matsuda, M.Hasegawa, N.Chikumoto, C.J. van der Beek and M. Konczykowski: J. Low Temp. Phys. {\bf 117} (1999) 1193; Y. Matsuda, A. Shibata, K. Izawa, H. Ikuta, M. Hasegawa and Y. Kato: Phys. Rev. B {\bf 66} (2002) 014527. 
\bibitem{UD} S. Ullah and A. T. Dorsey: Phys. Rev. B {\bf 44} (1991) 262. 
\bibitem{BK} See, for instance, T. R. Kirkpatrick and D. Belitz: Phys. Rev. Lett. {\bf 79} (1997) 3042. 
\bibitem{RI95} R. Ikeda: J. Phys. Soc. Jpn. {\bf 64} (1995) 1683. 
\bibitem{HN} B. I. Halperin and D. R. Nelson: J. Low Temp. Phys. {\bf 36} (1979) 599. The temperature $T_c^0$ in this reference corresponds to $T_{c0}$ in the present paper. 
\bibitem{Dorsey} R. J. Troy and A. T. Dorsey: Phys. Rev. B {\bf 47} (1993) 
2715. 
\bibitem{Maki} C-R. Hu: Phys. Rev. B {\bf 13} (1976) 4780 and references therein. 
\bibitem{RI8} R. Ikeda: J. Phys. Soc. Jpn. {\bf 66} (1997) 1603. 
\bibitem{com1} In contrast to a specific case (see Ref.23) of $s$-wave dirty films, the critical VG conductance value is {\it always} nonuniversal in general case where the parameters $\gamma_0$, $b_p$, and $b$ are independent of each other. 
\bibitem{PWA} P. W. Anderson: J. Phys. Chem. Solids {\bf 11} (1959) 26.  
\bibitem{Capan2} C. Capan, K. Behnia, Z. Z. Li, H. Raffy and C. Marin: Phys. Rev. B {\bf 67} (2003) 100507(R). 
\bibitem{Lee} See, for instance, P. A. Lee and M. G. Payne: Phys. Rev. B {\bf 5}(1972) 923. 
\bibitem{com4} Strictly speaking, the lower limit $0$ of the $s$-integral in the $\mu_n$-expression should be replaced by $T/\omega_c$, where $\omega_c$ is a high energy cutoff. The numerical $\rho(T)$ curves shown in the figures were unaffected by the choice of this lower limit in the $T/T_0$ ranges where we have numerically examined. 
\bibitem{com3} Although, in the $d_{x^2-y^2}$-pairing, cross terms between the lowest LL and the $n=4m$ ($m \geq 1$) higher LLs arise in the quadratic terms of eq.(2.1) so that the derivation of eq.(3.2) may be affected, the higher LLs can be  safely neglected unless $H/H_{c2}^*(0) \ll 1$.
\bibitem{NW} V. F. Mitrovic, H.N. Bachman, W.P. Halperin, A.P. Reyes, P. Kuhns and W.G. Moulton: Phys. Rev. B {\bf 66} (2002) 014511.
\bibitem{MIT} B. Khaykovich, Y.S. Lee, R.W. Erwin, S.-H. Lee, S. Wakimoto, K.J. Thomas, M.A. Kastner and R.J. Birgeneau: Phys. Rev. B {\bf 66} (2002) 014528.
\bibitem{Kivelson} S. A. Kivelson, D.-H. Lee, E. Fradkin and V. Oganesyan: Phys. Rev. B {\bf 66} (2002) 144516. 
\bibitem{Sachdev} E. Demler, S. Sachdev and Y. Zhang: Phys. Rev. Lett. {\bf 87} (2001) 067202. 
\bibitem{Ando} Y. Ando and K. Segawa: Phys. Rev. Lett. {\bf 88} (2002) 167005.
\bibitem{Loram} J. W. Loram, K. A. Mirza, J.R. Cooper and J.L. Tallon: Physica C {\bf 282-287}, 1405 (1997); N. Momono, T. Matsuzaki, M. Oda and M. Ido: J. Phys. Soc. Jpn. {\bf 71} (2002) 2832. 
\bibitem{Sasaki2} T. Sasaki, T. Fukuda, T. Nishizaki, T. Fujita, N. Yoneyama, N. Kobayashi and W. Biberacher: Phys. Rev. B {\bf 66} (2002) 224513. 
\bibitem{RI10} R. Ikeda: J. Phys.Soc.Jpn. {\bf 65} (1996) 33. 
\bibitem{Rochester} E. Bielejec and W. Wu: Phys. Rev. Lett. {\bf 88} (2002) 
206802. 
\bibitem{Phillips} P. Phillips and D. Dalidovich: Phil. Mag. {\bf 81} (2001) 
847. 
\bibitem{RI11} H. Ishida, H. Adachi and R. Ikeda: J. Phys. Soc. Jpn. {\bf 71} (2002) 245.  
\bibitem{RI12} H. Ishida and R. Ikeda: J. Phys. Soc. Jpn. {\bf 67} (1998) 983. 
\bibitem{Hebard} A. F. Hebard and M. A. Paalanen: Phys. Rev. Lett. {\bf 65} (1990) 927. 
\bibitem{Valles} J. A. Chervenak and J. M. Valles, Jr.: Phys. Rev. B {\bf 61} (2000) R9245. 
\bibitem{Gant} V. F. Gantmakher, M.V. Golubkov, V.T. Dolgopolov, G.E. Tsydynzhapov and A.A. Shashkin: JETP Lett. {\bf 71} (2000) 160. 
\bibitem{Goldman} N. Markovic, C. Christiansen, A.M. Mack, W.H. Huber and A.M. Goldman: Phys. Rev. B {\bf 60} (1999) 4320.
\bibitem{Pana} C. Panagopoulos, B.D. Rainford, J.R. Cooper, W. Lo, J.L. Tallon, J.W. Loram, J. Betouras, Y.S. Wang and C.W. Chu: Phys. Rev. B {\bf 60} (1999) 14617. 
\bibitem{Dima} V. B. Geshkenbein, L. B. Ioffe and A. J. Millis: Phys. Rev. Lett. {\bf 80} (1998) 5778. 
\bibitem{Lee2} P. A. Lee and G. Sha: cond-mat/0209572. 
\bibitem{Marcenat} C.Marcenat, S.Blanchard, J.Marcus, L.M.Paulius, C.J.van der Beek, M.Konczykowski and T. Klein: Phys. Rev. Lett. {\bf 90} (2003) 037004. 
\bibitem{Matsu} T. Nagaoka, Y. Matsuda, H. Obara, A. Sawa, T. Terashima, I. Chong, M. Takano and M. Suzuki: Phys. Rev. Lett. {\bf 80} (1998) 3594. 
\bibitem{Kokubo} N. Kokubo, J. Aarts and P. H. Kes: 
Phys. Rev. B {\bf 64} (2001) 014507. 
\bibitem{Ott} R. Jin and H. R. Ott: Phys. Rev. B {\bf 53} (1996) 9406. 
\bibitem{RI13} R. Ikeda: Physica C {\bf 316} (1999) 189. See the footnote of the first page there. 
\bibitem{Lang} W. Lang: private communication. 
\bibitem{Li} M. Li, C. J. van der Beek, M. Konczykowski, A.A. Menovsky and P.H. Kes: Phys. Rev. B {\bf 66} (2002) 024502. 
\bibitem{GaL1} V. M. Galitski and A. I. Larkin: Phys. Rev. B {\bf 63} (2001) 174506. 
\bibitem{GaL2} R. Ikeda: Phys. Rev. Lett. {\bf 89} (2002) 109703. 
\bibitem{Okuma} S. Okuma, S. Shinozaki and M. Morita: Phys. Rev. B {\bf 63} (2001) 054523.
\bibitem{LR} P. A. Lee and T. V. Ramakrishnan: Rev. Mod. Phys.{\bf 57} (1985) 287. 
\bibitem{Wang3} Y. Wang, S. Ono, Y. Onose, G. Gu, Y. Ando, Y. Tokura, 
S. Uchida and N. P. Ong: Science {\bf 299} (2003) 86. 
\bibitem{md} H. Balci, C. P. Hill, M. M. Qazilbash and R. L. Greene: cond-mat/0303469. 
\bibitem{Millis} For instance, see A. J. Millis: Phys. Rev. B {\bf 48} (1993) 7183. 

\end{thebibliography}
\end{document}